\documentclass[journal]{IEEEtran}
\usepackage{cite}
\usepackage[pdftex]{graphicx}
\usepackage[cmex10]{mathtools}
\usepackage{amssymb}
\usepackage{epsfig}
\usepackage{subfigure}
\usepackage{algorithm}
\usepackage{algpseudocode}
\usepackage[usenames,dvipsnames]{color}
\usepackage{array}

\usepackage{bm}
\usepackage{bbm}
\usepackage{microtype}

\newtheorem{lemma}{Lemma}
\newtheorem{prop}{Proposition}

\def\cA{{\mathcal{A}}} \def\cB{{\mathcal{B}}}  \def\cD{{\mathcal{D}}}
  \def\cG{{\mathcal{G}}}

\def\cQ{{\mathcal{Q}}}   
   
 \def\cZ{{\mathcal{Z}}} 

\def\ba{{\mathbf{a}}} \def\bb{{\mathbf{b}}} \def\bc{{\mathbf{c}}} \def\bd{{\mathbf{d}}}
\def\bee{{\mathbf{e}}} \def\bff{{\mathbf{f}}} \def\bg{{\mathbf{g}}} \def\bh{{\mathbf{h}}}

 \def\br{{\mathbf{r}}}  
\def\bu{{\mathbf{u}}} \def\bv{{\mathbf{v}}}

\def\bA{{\mathbf{A}}} \def\bB{{\mathbf{B}}} \def\bC{{\mathbf{C}}} \def\bD{{\mathbf{D}}}
 \def\bF{{\mathbf{F}}} \def\bG{{\mathbf{G}}} 
\def\bI{{\mathbf{I}}}

\DeclareMathOperator*{\argmin}{arg\,min}
\DeclareMathOperator*{\argmax}{arg\,max}

\begin{document}

%%%%%%%%%%%%%%%%%%%%%%%%%%%%%%%%%%%%%%%%%%%%%%
%%                                          %%
%%                  Title                   %%
%%                                          %%
%%%%%%%%%%%%%%%%%%%%%%%%%%%%%%%%%%%%%%%%%%%%%%

\title{Common Codebook Millimeter Wave Beam Design: Designing Beams for Both Sounding and Communication with Uniform Planar Arrays}

%%%%%%%%%%%%%%%%%%%%%%%%%%%%%%%%%%%%%%%%%%%%%%
%%                                          %%
%%                  Author                  %%
%%                                          %%
%%%%%%%%%%%%%%%%%%%%%%%%%%%%%%%%%%%%%%%%%%%%%%

\author{Jiho~Song,~\IEEEmembership{Student Member,~IEEE,} Junil~Choi,~\IEEEmembership{Member,~IEEE,} and David~J.~Love,~\IEEEmembership{Fellow,~IEEE}% <-this % stops a space
\thanks{J.\ Song and D.\ J.\ Love are with the School of Electrical and Computer Engineering, Purdue University, West Lafayette, IN 47907 (e-mail: \{jihosong, djlove\}@purdue.edu).}% <-this % stops a space
\thanks{{J.\ Choi is with the Department of Electrical Engineering, POSTECH, Pohang, Gyeongbuk 37673, Korea (e-mail:junil@postech.ac.kr).}}
\thanks{Parts of this paper were presented at the ICC, London, UK, June 8-12, 2015 \cite{Ref_Son14_2}.}
\thanks{This work was supported in part by the National Science Foundation (NSF) under grant CNS1642982  and the ICT R$\&$D program of MSIP/IITP. [2017(B0717-17-0002), Development of Integer-Forcing MIMO Transceivers for 5G $\&$ Beyond Mobile Communication Systems].}
}

% make the title area
\maketitle

%%%%%%%%%%%%%%%%%%%%%%%%%%%%%%%%%%%%%%%%%%%%%%
%%                                          %%
%%                 Abstract                 %%
%%                                          %%
%%%%%%%%%%%%%%%%%%%%%%%%%%%%%%%%%%%%%%%%%%%%%%

\begin{abstract}
Fifth generation (5G) wireless networks are expected to utilize wide bandwidths available at millimeter wave (mmWave) frequencies for enhancing system throughput. However, the unfavorable channel conditions of mmWave links, e.g., higher path loss and attenuation due to atmospheric gases or water vapor, hinder reliable communications. To compensate for these severe losses, it is essential to have a multitude of antennas to generate sharp and strong beams for directional transmission. In this paper, we consider mmWave systems using uniform planar array (UPA) antennas, which effectively place more antennas on a two-dimensional grid. A hybrid beamforming setup is also considered to generate beams by combining a multitude of antennas using only a few radio frequency chains. We focus on designing a set of transmit beamformers generating beams adapted to the directional characteristics of mmWave links assuming a UPA and hybrid beamforming. We first  define ideal beam patterns for UPA structures. Each beamformer is constructed to minimize the {mean squared error} from the corresponding ideal beam pattern. Simulation results verify that the proposed codebooks enhance beamforming reliability and data rate in mmWave systems.
\end{abstract}

\begin{IEEEkeywords}
Millimeter wave communications, Hybrid beamforming, Codebook design algorithm,  Uniform planar array.
\end{IEEEkeywords}

\IEEEpeerreviewmaketitle

\section{Introduction}
\IEEEPARstart{W}{ireless} broadband systems operating in the millimeter wave (mmWave) spectrum are thought to be
a prime candidate to provide the system throughput enhancements needed for fifth generation (5G) wireless  networks \cite{Ref_Gho14,Ref_Rap13,Ref_Pi11,Ref_Roh14}. The wide bandwidths available at mmWave frequencies can be an attractive alternative to the sub-$6$GHz frequencies employed in most of today's cellular networks. Also, the directional characteristics of mmWave links are suitable for reducing interuser interference in multiuser channels. However, the higher expected path loss caused by the high carrier frequency, atmospheric gases, and water vapor absorption result in {severe link quality degradation.} The unfavorable channel conditions at mmWave frequencies necessitate utilizing highly directional transmission with a large beamforming gain.

{The small wavelengths of mmWave frequencies {allow} a large number of antennas to be implemented in a small form factor {on access points and devices. Phased} array transmit/receive {architectures, such as a uniform linear array (ULA) or uniform planar array (UPA), using high-resolution beamforming are usually considered for mmWave systems \cite{Ref_Han98}.} Due to their simplicity, systems using ULAs have been widely studied for use at mmWave frequencies \cite{Ref_Hur11,Ref_Hur13,Ref_Son13,Ref_Son14,Ref_IEEE802.11,Ref_Sun14,Ref_Noh15,Ref_Alk14,Ref_Aya14}. {UPAs are now} being considered due to their higher space efficiency, obtained by packing antennas on a two-dimensional (2D) grid \cite{Ref_Noh15}. UPAs can also facilitate three-dimensional (3D) beamforming that  takes advantage of both elevation   and azimuth  domain beamforming to efficiently mitigate interuser interference and eventually increases  system capacity \cite{Ref_Nam13}. In this paper, we thus consider UPAs to take the advantage of the 2D antenna structures.}

Millimeter wave systems having a large number of antennas may not be able to use baseband beamforming techniques requiring one radio frequency (RF) chain per antenna due to the high cost and power consumption \cite{Ref_Pi11, Ref_Roh14}. Therefore, mmWave systems based on analog beamforming relying upon a single RF chain have been reported in \cite{Ref_Hur11,Ref_Hur13,Ref_Son13,Ref_Son14,Ref_IEEE802.11,Ref_Sun14}.  {Although {an analog beamforming architecture can be implemented with inexpensive transmit amplifiers, the envelope constraints} placed on the transmitted signals may result {in a loss} of beamforming gain \cite{Ref_Lov03,Ref_Hur13}. To develop mmWave systems operating under modest hardware requirements, we  consider hybrid beamforming techniques using a few RF chains wired to sets of phase shifters \cite{Ref_Sun14,Ref_Alk14,Ref_Aya14,Ref_Noh15}. }

Current cellular systems {construct the transmit beamformer} based on the channel state information (CSI) at the transmitter, which often is only available through receiver feedback. In feedback-assisted frequency division duplexing architectures,  it is essential for the receiver to estimate the CSI using downlink pilot signals {\cite{Ref_Lov08,Ref_Has03,Ref_San10}.} In large-scale mmWave systems, it may be difficult to explicitly estimate the CSI due to the large number of resources {required for training each} antenna \cite{Ref_Has03,Ref_San10}. Channel estimation algorithms relying on compressed sensing techniques \cite{Ref_Alk14,Ref_Bar10,Ref_Ram12,Ref_Kim15} could be suitable for mmWave downlink training. {However, compressed sensing techniques need stringent sparsity requirements to be satisfied.

{The high dimensionality of a mmWave channel {necessitates}  utilizing a more intuitive channel estimation algorithm. For this reason, mmWave systems may use beam alignment approaches that choose the  transmit beamformer without estimating the channel matrix explicitly \cite{Ref_Hur11,Ref_Hur13,Ref_Son13,Ref_Son14,Ref_IEEE802.11}. In codebook-based beam alignment approaches, it is desirable to share a single \textit{common} codebook for both channel sounding and data transmission.} However, {sounding and data transmission enforce} conflicting design requirements on the beams in the codebook.  For example, data transmission beams {should} ideally be narrow to allow for maximum beamforming gain when properly aligned, but channel sounding beams {should} ideally be {wide} to sound a wide geographic {area with small overhead.}  In addition, the codebook size must be small to ensure minimal system overhead.

{The design of a common codebook  satisfying the conflicting design requirements as well as validating practical mmWave systems  has been studied  in \cite{Ref_Hur13,Ref_Son14_2,Ref_Alk14}.} Previous codebook design algorithms are typically optimized for  particular vector subspaces characterized by ULA structures. However, codebooks for mmWave systems employing UPAs should be designed to sound a  wide geographic area as well as to facilitate a large beamforming gain. In addition, adaptive beam alignment approaches mostly utilize a multitude of hierarchical codebooks  \cite{Ref_Hur13,Ref_Son14,Ref_Noh15,Ref_Alk14}. This necessitates design guidelines for multi-resolution codebooks {that capture the channel characteristics of UPAs.}

\begin{figure}[!t]
\normalsize
\centering
{\includegraphics[width=0.48\textwidth]{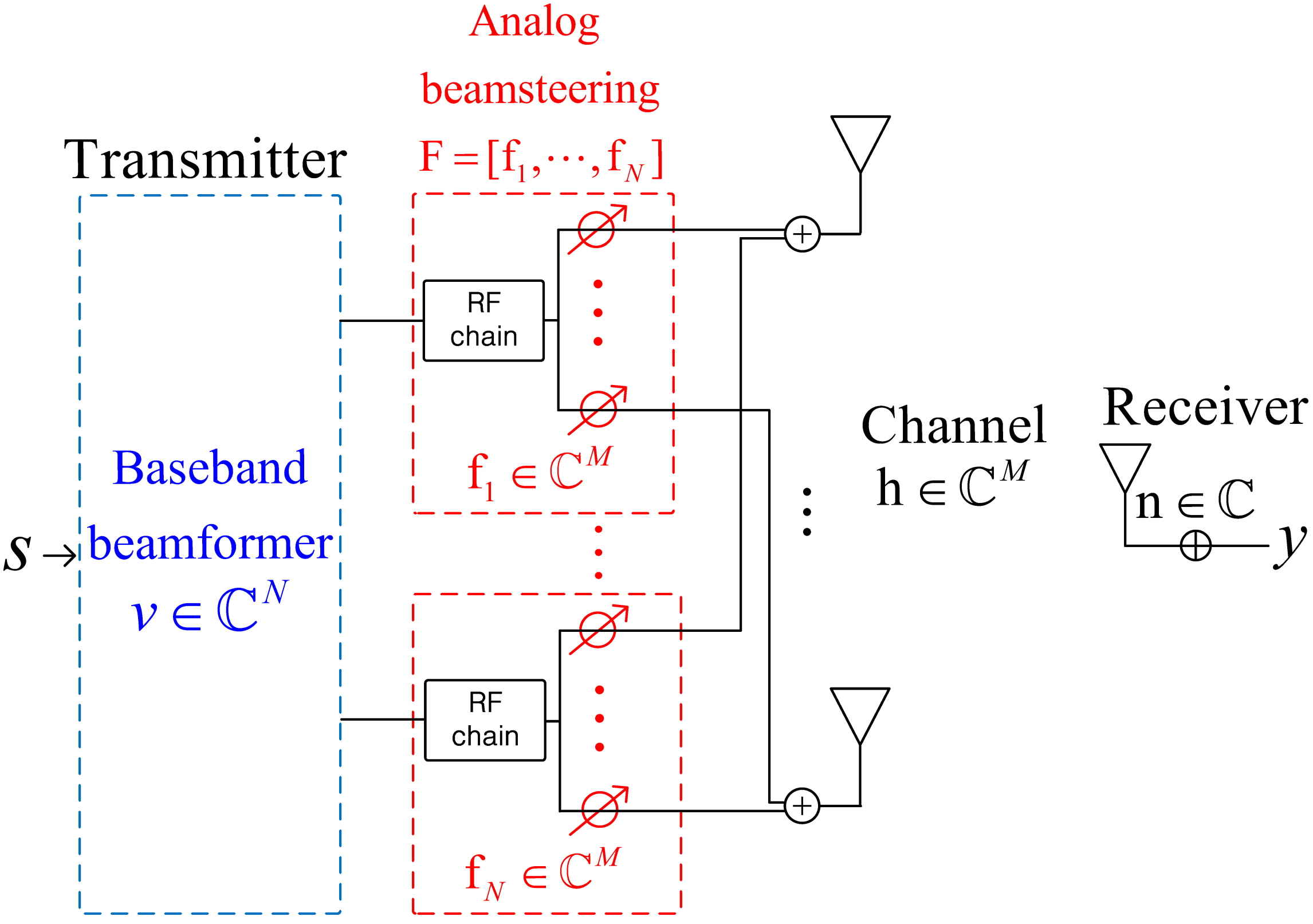}}
\caption{An overview of a mmWave system with hybrid beamforming.}
\label{fig:system_overview}
\end{figure}

{In this paper, we propose a practical codebook design algorithm that utilizes the strong directivity of mmWave links. The codebook design algorithm is developed assuming hybrid beamforming at the transmitter. Based on Parseval's theorem, we first derive conditions that impose constraint on  a beamformer's beam pattern assuming a UPA structure. To develop a codebook design criterion, we next {study ideal beam patterns}  by utilizing the analytical studies. The codebook is {designed such that each beamformer} minimizes the mean squared error (MSE) between the codebook's beam pattern and the corresponding ideal beam pattern. To access a feasible solution satisfying the proposed design guideline, we formulate an optimization problem that can produce a set of  candidate beamformers. The orthogonal matching pursuit (OMP) algorithm \cite{Ref_Tro07, Ref_Reb02} is then used to compute each candidate beamformer, satisfying a power constrained hybrid beamforming setup. The final beamformer accomplishing the MSE minimization objective will be chosen among the set of beamformer candidates for each ideal beam pattern. It is validated analytically that all other beamformers in the codebook can be generated from one optimized beamformer, which can expedite offline codebook construction.} {Although the ULA codebooks in \cite{Ref_Hur13,Ref_Alk14} can be extended to UPAs similar to the 2D Kronecker product (KP) codebook in \cite{Ref_3GPP150560}, the simulation results and discussions in Section \ref{sec:SR} verify {that the proposed codebooks} generate the most practical beam patterns, which are suitable to  the codebook-based beam alignment approaches.}

The remainder of this paper is organized as follows. In Section II, we describe a mmWave system with hybrid beamforming and {define the {beam region of interest  (beam region)}. In Section III, we define the ideal beam pattern} by considering a UPA structure and the directional characteristics of mmWave links. In Section IV, a practical codebook design algorithm is proposed for mmWave systems {based on predefined ideal beam patterns.} In Section V, simulation results are presented to verify the performance of the proposed codebook. Section VI details our conclusions.

Throughout this paper, $\mathbb{C}$ denotes the field of complex numbers, $\mathbb{R}$ denotes the field of real numbers, $\mathcal{CN}(m,\sigma^2)$ denotes the complex normal distribution with mean $m$ and variance $\sigma^2$, $[a,b]$ is the closed interval between $a$ and $b$, $\mathrm{U}(a,b)$ denotes the uniform distribution in the closed interval $[a,b]$, $(\ba)_{\ell}$ is the $\ell$-th entry of the column vector $\ba$, {$\mathbf{1}_{a,b}$ is the $a \times b$ all ones matrix}, $\mathbf{I}_{N}$ is the $N \times N$ identity matrix,  $\lceil~\rceil$ is the ceiling function, $\mathrm{E}[\cdot]$ is the expectation operator, $\mathbbm{1}$ is the indicator function,  $\| \cdot \|_p$ is the $p$-norm, $\odot$ is the Hadamard product, and $\otimes$ is the Kronecker product. Also, $\bA^H$, $\bA^*$, $\bA_{a,b}$, $\bA_{a,:}$, $\bA_{:,b}$, $\mathfrak{v}_{\textrm{max}}\{\bA\}$ denote the conjugate transpose, element-wise complex conjugate, $(a,b)^{th}$ entry, $a^{th}$ row, $b^{th}$ column, and principal eigenvector of the matrix $\bA$, respectively.

\begin{figure}[!t]
\centering
{\includegraphics[width=0.31\textwidth]{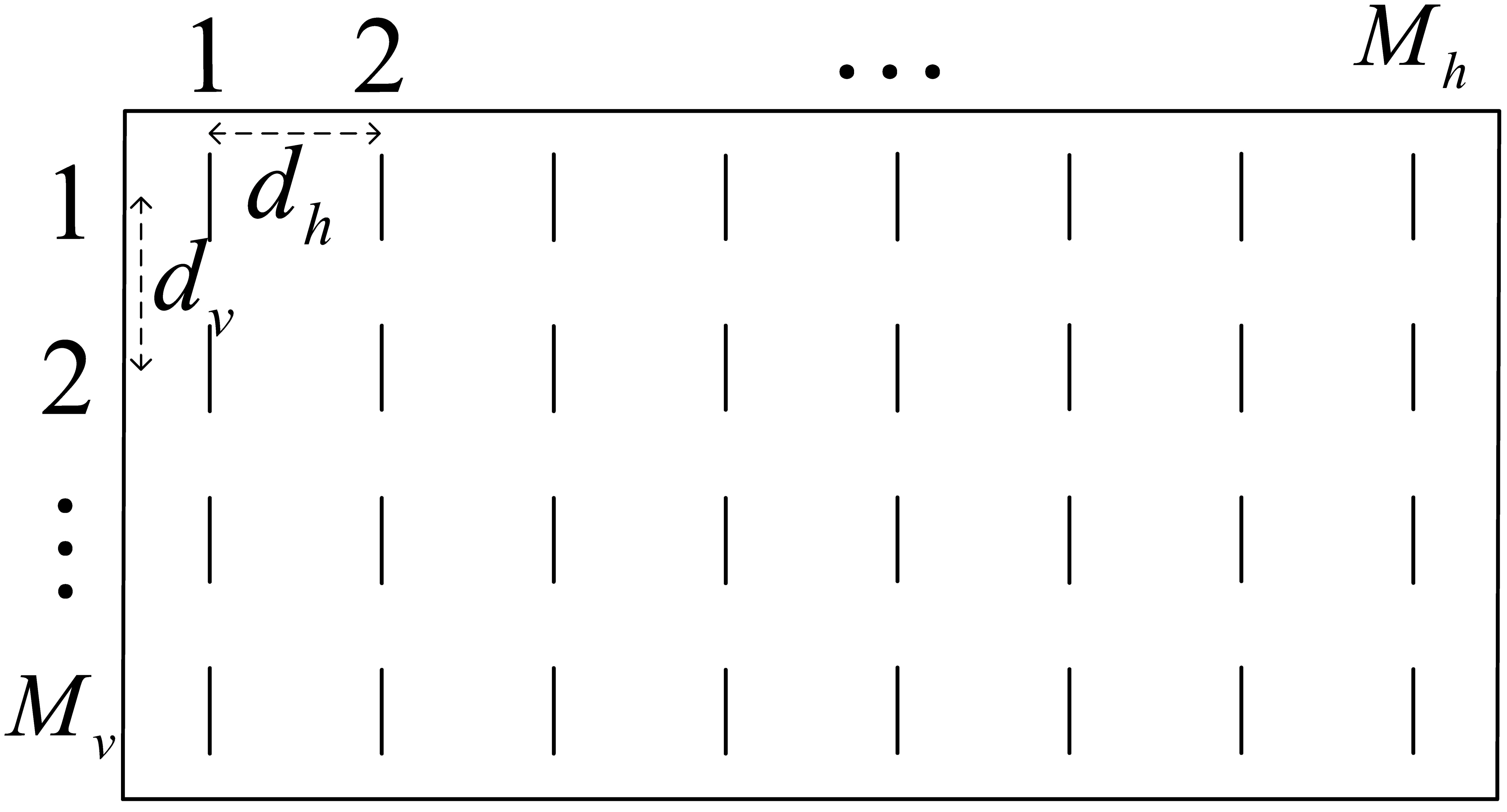}}
\caption{{The structure of the uniform planar array considered in this paper.}}
\label{fig:upa}
\end{figure}

\section{System Model}
\subsection{{System model}}
We consider a multiple-input single-output (MISO) system\footnote{{For multiple-input multiple-output (MIMO) systems, we also need to perform beam alignment at the receiver as in \cite{Ref_Hur11,Ref_Hur13}. Although we  discuss only beamformer design at the transmitter for simplicity, the proposed codebook design algorithm can be used to construct combiners at the receiver as well.}} operating in the mmWave spectrum. The transmitter employs $M \doteq M_h  M_v$ transmit antennas, which are controlled by $N$ RF chains ($N \leq M$), and the receiver has a single receive antenna \cite{Ref_Alk14,Ref_Aya14}. This hybrid beamforming configuration is shown in Fig. \ref{fig:system_overview}. The transmit array is laid out in a grid pattern with $M_h$ columns and $M_v$ rows, as shown in Fig. \ref{fig:upa}.  The horizonal and vertical elements are spaced uniformly with separations $d_h$ and $d_v$, respectively \cite{Ref_Han98}.

Assuming a block fading channel, the input-output expression for data transmission is
\begin{align}
\label{eq:sample}
y \doteq \sqrt{\rho}\bh^H\bc s+n,
\end{align}
where $y$ is the received signal, $\rho$ is the transmit signal-to-noise ratio (SNR), {$\bc \in \mathbb{C}^{M}$ is the unit norm transmit beamformer} $\bh \in \mathbb{C}^{M}$ is the  block fading mmWave channel, $s \in \mathbb{C}$ is the transmit symbol subject to the constraint $\mathrm{E}[|s|^2]\leq 1$, and {$n \sim \mathcal{CN}(0,1)$ is the additive white Gaussian noise (AWGN).}

{In most beam alignment approaches, the transmit beamformer for data transmission  is chosen from a sounding codebook $\mathcal{C}=\big\{\bc_{1,1}\cdots\bc_{Q_h,Q_v} \big\}$ consisting of $Q\doteq Q_h Q_v$ beamformers.\footnote{Because it is possible to easily deploy a large number of transmit antennas, we assume that $Q < M$ in order to ensure minimal system overhead for beam alignment.}  {To sound the mmWave channel, the codewords $\{\bc_{1,1}\cdots\bc_{Q_h,Q_v}  \}$ are transmitted one-by-one such as
\begin{align*}
y_{q,p}^{{\mathrm{sounding}}}= \sqrt{\rho}\bh^H\bc_{q,p}+n_{q,p},
\end{align*}
where $q\in\{1,...Q_h\}$, $p\in\{1,...,Q_v\}$ Note that $y_{q,p}^{{\mathrm{sounding}}}$ is the $((q-1)Q_v+p)$-th channel observation in the given fading block and $n_{q,p} \sim \mathcal{CN}(0,1)$ is  AWGN. The beamformer for data transmission is then given by  $\bc =\bc_{\hat{q},\hat{p}}$  {with}
\begin{align}
\label{eq:beam}
(\hat{q},\hat{p})&=\argmax_{(q,p)  \in \cQ  }   { \big|  y_{q,p}^{{\mathrm{sounding}}} \big|^2},
\end{align}
where $\cQ \doteq \{1, \cdots, Q_h\} \times \{1, \cdots, Q_v\}$.} % Note that the variable $i$ denoting the channel use is dropped for simplicity.

{When implemented,} the unit norm  transmit beamformer $\bc\doteq\bF\bv \in \mathbb{C}^{M}$ is formed using a combination of an analog beamsteering  matrix $\bF=\big[\bff_{1},\cdots,\bff_{N}\big] \in \mathbb{C}^{M \times N}$ consisting of $N$ unit norm beamsteering vectors and a baseband beamformer {$\bv \in \mathbb{C}^{N}$}. {An analog beamsteering vector $\bff_{n}$ is realized by a set of RF phase shifters, {which is modeled by requiring that the vector lie in the} {equal gain subset
\begin{align}
\label{eq:equal_set1}
\mathcal{E}_{2^B} &= \big\{\bff \in \mathbb{C}^{M} : f_{m} = e^{j \varphi_m}/\sqrt{M},~\varphi_m \in \mathcal{Z}_{2^B} \big\},
\end{align}
where $m \in \{ 1,\cdots,M\}$, $\varphi_m$ is the phase of the each entry of the equal gain vector $\bff=[f_1,\cdots, f_M]^T$.} To impose practical limitations on the analog beamforming hardware, we assume that a digitally-controlled RF phase shifter generates  quantized phases \cite{Ref_Hur13}. Each element phase $\varphi_m$ in (\ref{eq:equal_set1}) is then chosen from the set of  $2^B$  quantized phases}
\begin{align}
\label{eq:phase}
\mathcal{Z}_{2^B} \doteq \big\{0,{2\pi}/{2^B},\cdots,{2\pi(2^B-1)}/{2^B} \big\}.
\end{align}
The weight vector $\bv$ {combining the columns of $\bF$} performs beamforming at baseband without equal gain constraints, while the combination of $\bF$ and $\bv$ is subject to the constraint $\|\bF\bv \|_2^2=1$.

\subsection{Vector subspace for mmWave channels}

For the UPA scenario, the mmWave channel is modeled by the combination of a line-of-sight (LOS) path and a few {non-line-of-sight (NLOS)} paths as \cite{Ref_Rap12,Ref_Akd14,Ref_Zha10}
\begin{align}
\nonumber
\bh &=  \sqrt{\frac{MK}{1+K}} \alpha_0  \mathbf{d}_{M}(\psi_{h0},\psi_{v0})
\\
\label{eq:channel}
&~~~~~~~~~~~~~~~~~~~+\sqrt{\frac{M}{R(1+K)}} \sum_{r=1}^{R} \alpha_r \mathbf{d}_{M}(\psi_{hr},\psi_{vr}),
\end{align}
where $K$ is the Ricean $K$-factor, $\alpha_r \sim \mathcal{CN}(0,1) $ is the complex channel gain, $R$ is the number of NLOS paths, and
\begin{align}
\label{eq:ray}
\mathbf{d}_{M}(\psi_{hr},\psi_{vr}) \doteq \mathbf{d}_{M_h}(\psi_{hr}) \otimes \mathbf{d}_{M_v}(\psi_{vr}) \in \mathbb{C}^{M}
\end{align}
is the $r$-th normalized beam defined by the KP of array response vectors\footnote{{For simplicity, we do not consider the electromagnetic interaction between deployed antennas \cite{Ref_Cle07}. If we define the effective array response vector  taking the electromagnetic mutual couplings into account,  the proposed approach can be directly applied to generate practical beam patterns that consider the mutual coupling between deployed antennas.}}
\begin{align}
\label{eq:array_vector}
&\mathbf{d}_{M_{a}}(\psi_{ar}) \doteq  \frac{1}{\sqrt{M_a}} \big[1,e^{j\psi_{ar}}\cdots e^{j ({M_a}-1)\psi_{ar}} \big]^{T} \in \mathbb{C}^{M_a}
\end{align}
{with  $\psi_{hr}=\frac{2\pi d_{h}}{\lambda}\sin \theta_{hr}\cos \theta_{vr}$ and $\psi_{vr}=\frac{2\pi d_{v}}{\lambda}\sin \theta_{vr}$, where  $\theta_{ar}$ is the angle of departure (AoD) \cite{Ref_Han98}. Note that $a \in \{h,v \}$ {denotes} both horizontal and vertical domains.}

{Millimeter wave channels are expected to have a large Ricean $K$-factor that is matched with a strong channel directivity \cite{Ref_Muh10,Ref_Pi11,Ref_Say13}. Therefore, we consider a vector subspace defined by a single dominant beam,\footnote{{Although we mainly focus on designing a codebook for a single dominant beam, we also consider channels consisting of multiple NLOS paths for numerical simulations.}} i.e., an array manifold
\begin{align}
\label{eq:am}
\cA \doteq \big\{\ba : \ba=\sqrt{M}\mathbf{d}_{M}(\psi_{h},\psi_{v}),~ (\psi_{h},\psi_{v})  \in \cB \big\},
\end{align}
where $\cB$ denotes the set of beam directions\footnote{We call the 2D geometric model of the vector subspace as a {beam region}.} in both horizontal and vertical  domains. {Then, we need to define proper $\mathcal{B}$ to design good codebooks.}}

{{Since the array response vector is periodic such as
\begin{align*}
\bd_{M_a}(\psi_a+2\pi)=\bd_{M_a}(\psi_a),
\end{align*}
the entire {beam region}  is bounded as $\cB_e \doteq  [-\pi,\pi) \times [-\pi,\pi)$.} We next define the set of possible beam directions that actually characterizes the array manifold in (\ref{eq:am}).  We consider an AoD distributed as $(\theta_h,\theta_v) \in [-\frac{\pi}{2},\frac{\pi}{2}) \times  [-\frac{\pi}{4},\frac{\pi}{4})$ assuming sectorized cellular systems. The {beam region}  is then defined as
\begin{align*}
\cB_{s} &\doteq \big[-2\pi d_h/\lambda ,2\pi d_h /\lambda \big) \times  \big[-\sqrt{2}\pi d_v /\lambda,\sqrt{2}\pi d_v/\lambda \big).
\end{align*}
As defined in (\ref{eq:array_vector}), the beam direction in $\psi_h$ domain is a function of not only $\theta_h$ but also $\theta_v$. Considering the paired ranges in the horizontal and vertical domains together, the possible range of beam directions are bounded for a given AoD $\theta_v$ {as}
\begin{align*}
|\psi_{h|\theta_v}| \leq 2\pi d_h  \cos\theta_{v} / \lambda,~~~\psi_{v|\theta_v} = 2\pi d_v  \sin \theta_{v} / \lambda.
\end{align*}
{Under the assumption of {$d_h=d_v=d$,} the bounds in both domains give}
\begin{align}
\label{eq:cir}
\psi_{h|\theta_v}^2+\psi_{v|\theta_v}^2 \leq ({2\pi d}/{\lambda})^2.
\end{align}
The  {beam region} is thus defined by  considering a feasible set of beam directions,\footnote{In case of $d_h \ne d_v$, the {beam region} is defined by an ellipse such as $\cB_{c} \doteq \Big\{ (\psi_{h},\psi_{v}) \in \cB_s : \frac{\psi_{h}^2}{({2\pi d_h}/{\lambda})^2} + \frac{\psi_{v}^2}{({2\pi d_v}/{\lambda})^2} \leq 1 \Big\}$.} which satisfy the condition in (\ref{eq:cir}), over all possible AoDs $\theta_v$}
\begin{align*}
\cB_{c} &\doteq  \big\{ (\psi_{h},\psi_{v}) \in \cB_s: \psi_{h}^2 + \psi_{v}^2 \leq ({2\pi d}/{\lambda})^2 \big\}.
\end{align*}

\begin{figure}[!t]
\centering
{\includegraphics[width=0.415\textwidth]{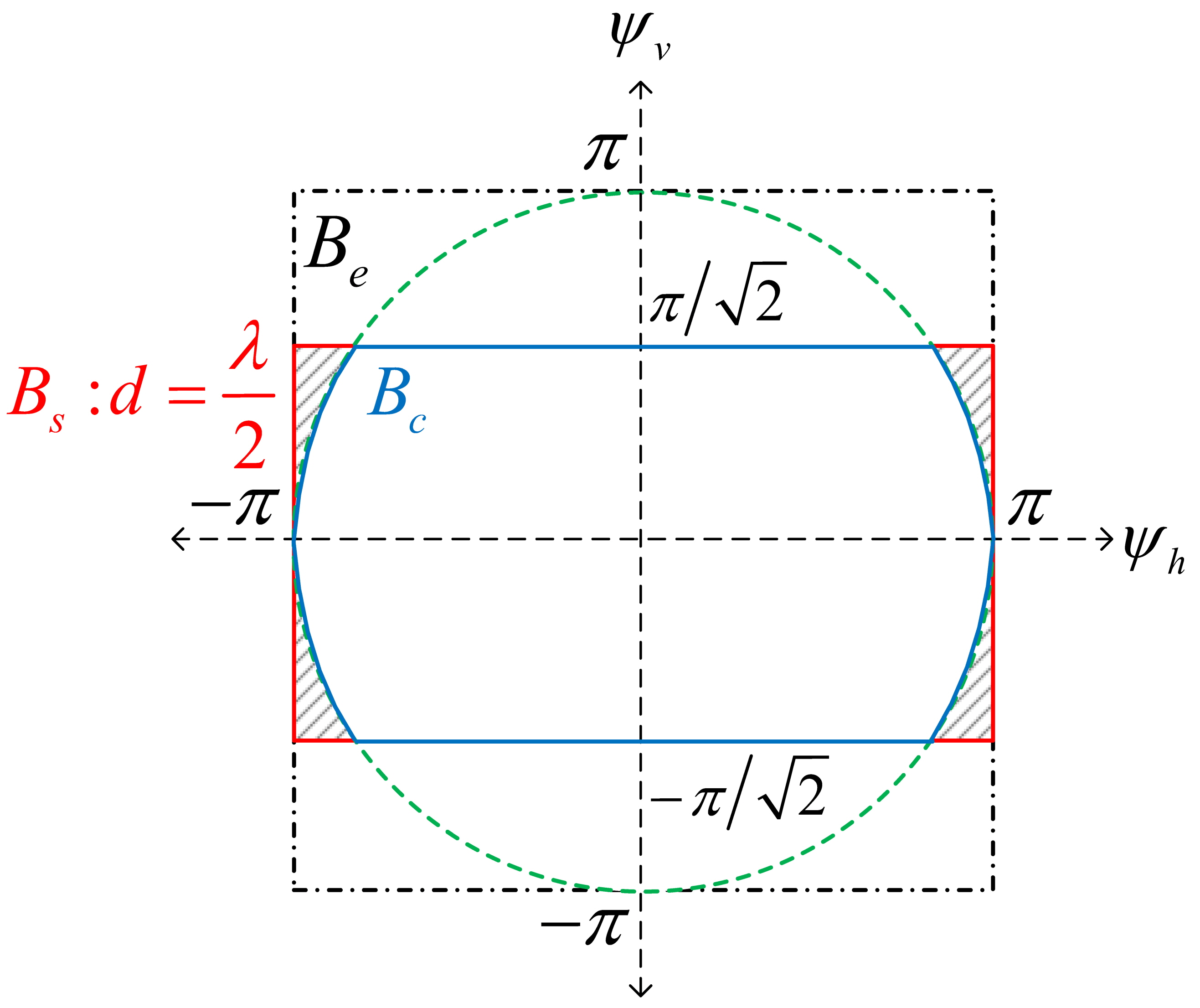}}
\caption{Two-dimensional {beam region} under UPA scenario.}
\label{fig:bs_all}
\end{figure}

{We now take a closer look at the  {beam region} $\cB_{c}$. In Fig. \ref{fig:bs_all}, the patterned areas $ \cB_{s} \backslash \cB_{c}$ excluded from $\cB_{c}$ (which describes the {pairing effect} between the horizontal and vertical domains) are negligible compared to the main area. For simplicity, we develop a codebook design algorithm without taking into account the {pairing}  between horizontal and vertical domains. For the assumption of  $d=\lambda/2$, the beam directions {are simplified as $\psi_a=\pi \sin \theta_a$, and the {beam region} is bounded  as}
\begin{align}
\label{eq:set_ac}
\cB_s  \doteq \big[ -{\psi}^{\textrm{B}}_h , {\psi}^{\textrm{B}}_h \big) \times   \big[ -{\psi}^{\textrm{B}}_v,{\psi}^{\textrm{B}}_v\big)
\end{align}
with  ${\psi}^{\textrm{B}}_h=\pi$,  and ${\psi}^{\textrm{B}}_v=\pi/\sqrt{2}$. Finally, we define the array manifold {by setting $\mathcal{B}=\mathcal{B}_s$ in (\ref{eq:am}).}}

\section{Framework for codebook design algorithm}

{To use a common codebook for both channel sounding and data transmission, beamformers should be designed to generate beam patterns satisfying the following criteria:}
\\
{\textbf{1)} Each beamformer should cover a wide geographic area to reduce sounding overhead.}
\\
{\textbf{2)} Each beamformer should have high and uniform gain {over} a desired {beam region} to maximize SNR and quality of service.}

{To provide design guidelines for practical beamformers that satisfy the above criteria, {it is necessary to {design} beam patterns having uniform beamforming gains inside the {beam region} (to meet the quality of service requirement) as well as having no beamforming gain outside the {beam region (to maximize SNR, minimize beam misalignment, and reduce interference)}.} We will refer to these beam patterns as \textit{ideal beam patterns} throughout the paper.\footnote{{The term \textit{ideal} does not mean the considered beam patters are globally optimal beam patterns.}}}

\begin{figure}[!t]
\centering
{\includegraphics[width=0.49\textwidth]{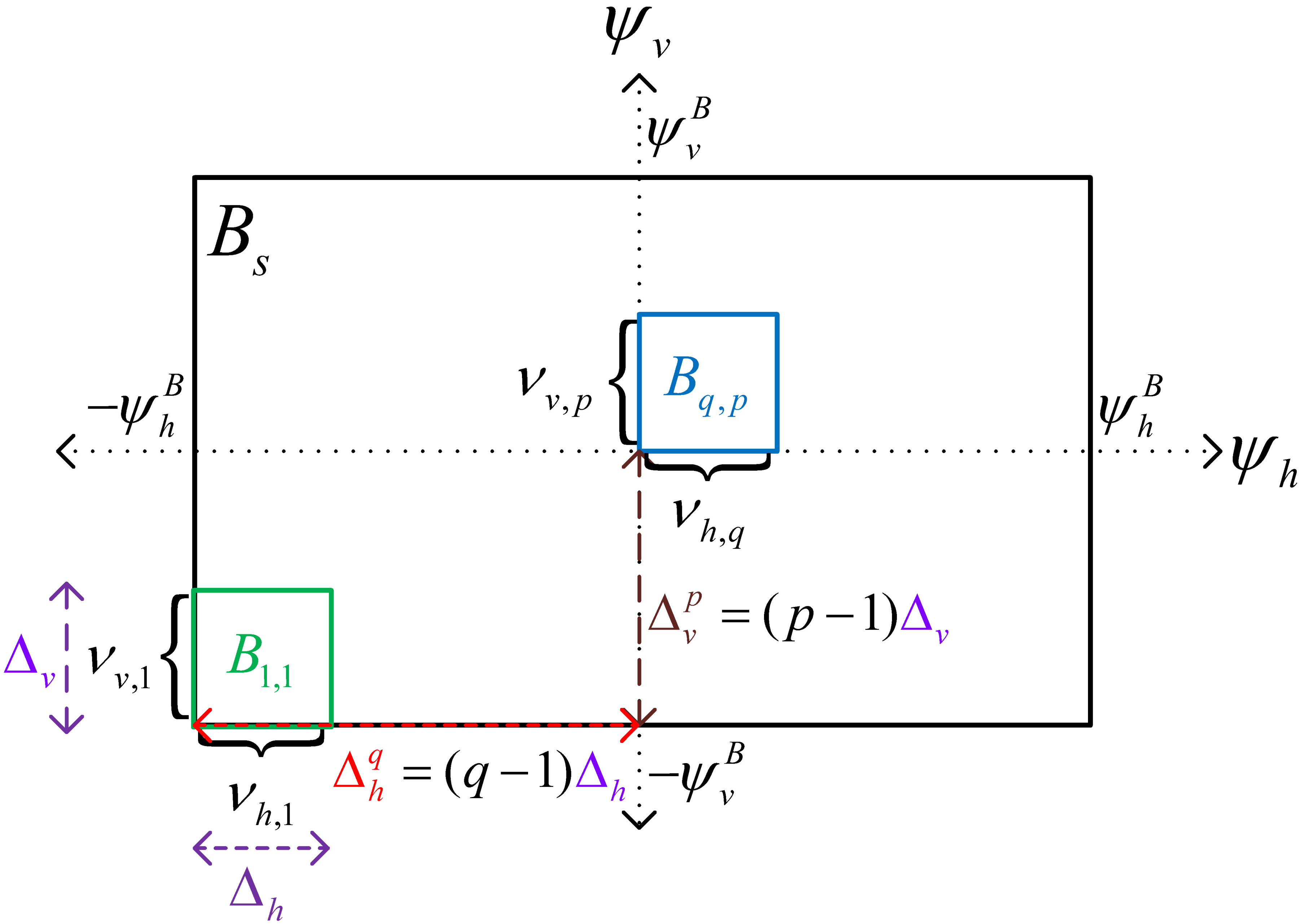}}
\caption{{Target {region of interest} for each beamformer.}}
\label{fig:subspace}
\end{figure}

{Assuming a uniform distribution of users, we equally divide the {beam region}  $\cB_s$ in (\ref{eq:set_ac}) into $Q=Q_hQ_v$ subspaces. To offer the same quality of service over the coverage region, each beamformer is designed to generate a beam pattern that covers its target {beam region}. As depicted in Fig. \ref{fig:subspace}, {the beam region} defined for the corresponding beamformer $\bc_{q,p}$ is  formed by the combination  of the $q$-th and $p$-th ranges {in the horizontal and vertical domains as}
\begin{align}
\label{eq:BS}
\cB_{q,p}& \doteq \nu_{h,q} \times \nu_{v,p}
\end{align}
where the $b$-th range in the domain $a \in \{h,v\}$ is defined as
\begin{align}
\label{eq:ebw}
\nu_{a,b} & \doteq - {\psi}^{\textrm{B}}_a +  \Delta_a^b+  [0,\Delta_a ).
\end{align}
Note that $\Delta_a^b \doteq (b-1)\Delta_a$ is the shifted beam direction, and $\Delta_a  \doteq   2\psi^{\textrm{B}}_a/{Q_a}$ is the beam-width of a beamformer.}

Each beamformer  satisfying the second criterion  has to generate higher reference gains  for its target {beam region}  as well as generate smaller reference gains for the rest of the {beam region}. Note that the reference gain  is defined as
\begin{align*}
G(\psi_h,\psi_v,\bc) \doteq \big| \big(\mathbf{d}_{M_h}(\psi_h) \otimes \mathbf{d}_{M_v}(\psi_v) \big)^H\bc \big|^2.
\end{align*}

Before developing the ideal, but unachievable in practice, beam patterns, we first discuss a constraint condition that is subject to any beam patterns in the following lemma.
\begin{lemma}
\label{lm:01}
{The integral with respect to the  beam pattern over $\cB_e = [-\pi,\pi)\times [-\pi,\pi)$ is
\begin{align*}
\int_{-\pi}^{\pi}  \int_{-\pi}^{\pi} G(\psi_h,\psi_v,\bc)  d\psi_h  d\psi_v=\frac{(2 \pi)^2}{M}
\end{align*}
for any unit norm vector $\bc \in \mathbb{C}^M$.}
\end{lemma}
\begin{IEEEproof}
The integration of the reference gain is
\begin{align*}
&\int_{-\pi}^{\pi}  \int_{-\pi}^{\pi} G(\psi_h,\psi_v,\bc)  d\psi_h  d\psi_v
\\
& \stackrel{(a)} = \int_{-\pi}^{\pi}  \int_{-\pi}^{\pi}  \bigg| \sum_{\ell=1}^{M_h} e^{j(1-\ell)\psi_h} \mathbf{d}_{M_v}^H(\psi_v) \bc_{\ell}        \bigg|^2       \frac{d\psi_h d\psi_v}{M_h}
\\
& = \int_{-\pi}^{\pi}  \int_{-\pi}^{\pi}  \bigg|   \sum_{\ell=1}^{M_h}  e^{j(1-\ell)\psi_h} \sum_{m=1}^{M_v}     e^{j(1-m)\psi_v} (\bc_{\ell})_{m} \bigg|^2     \frac{d\psi_hd\psi_v}{M_h M_v}
\\
& \stackrel{(b)} = \frac{2\pi}{M} \sum_{\ell=1}^{M_h}  \int_{-\pi}^{\pi}    \bigg| \sum_{m=1}^{M_v}  e^{j(1-m)\psi_v} (\bc_{\ell})_{m} \bigg|^2    d\psi_v
\\
& \stackrel{(c)} = \frac{(2\pi)^2}{M} \sum_{\ell=1}^{M_h}\sum_{m=1}^{M_v}  | (\bc_{\ell})_{m} |^2
\\
&=  \frac{(2 \pi)^2}{M},
\end{align*}
where $(a)$ is derived because $\bc \doteq [\bc_1^T,\cdots,\bc_{M_h}^T]^T,~\bc_{\ell} \in \mathbb{C}^{M_v}$ and {$(b)$, $(c)$ are derived based on the Parseval's theorem \cite{Ref_Hur13}
\begin{align*}
\frac{1}{2 \pi}\int_{-\pi}^{\pi} \bigg| \sum_{m=1}^{M}  e^{j(m-1)\psi}(\ba)_{m} \bigg|^2 d\psi = \| \ba \|_2^2
\end{align*}
for any vector $\ba \in \mathbb{C}^M$.}
\end{IEEEproof}

We now define \textit{ideal beam patterns}  by taking Lemma \ref{lm:01} into account. We assume the $(q,p)$-th ideal beamformer generates an ideal beam pattern that is biased toward its desired {beam region}. For example, the $(q,p)$-th beamformer generates an ideal beam pattern  having a non-zero reference gain for the {beam region} $\cB_{q,p}$ and zero gain for the rest of the {beam region} $\cB_s  \setminus \cB_{q,p}$.  {The ideal beam pattern is then given by
\begin{align*}
 G_{q,p}^{\textrm{ideal}}(\psi_h,\psi_v)&=t(\psi_{h},\psi_{v})\mathbbm{1}_{\cB_{q,p}}(\psi_h,\psi_v)
\end{align*}
subject to the constraint condition derived in Lemma \ref{lm:01} making}
\begin{align}
\label{eq:con}
\int \int_{\mathcal{B}_{q,p}}t(\psi_{h},\psi_{v})d\psi_h  d\psi_v=\frac{(2 \pi)^2}{M}.
\end{align}

{We discuss a distribution of the non-zero reference gain $t(\psi_{h},\psi_{v})$. Assuming a single  dominant beam $\sqrt{M} \mathbf{d}_{M}(\psi_{h},\psi_{v})$ is in $\cB_{q,p}$, the $(q,p)$-th ideal beamformer is an optimal transmit beamformer. Under the assumption of a LOS dominant channel model, we define  the expected data rate conditioned on $\|\bh \|_2^2$ as
\begin{align*}
 \mathrm{R} &\doteq \mathrm{E}\Big[  \log_{2} \big( 1+ \rho | \bh^H \bc_{q,p} |^2 \big) ~\big|~ \| \bh \|_2^2 \Big]
 \\
 &=\mathrm{E}\Big[  \log_{2} \big({1+ \rho \|\bh \|_2^2  |\mathbf{d}_{M}^H(\psi_{h},\psi_{v})\bc_{q,p}  |^2 }\big) ~\big|~ \| \bh \|_2^2 \Big].
\end{align*}
In the following lemma, we define the non-zero reference gains  that achieve an upper bound of the expected data rate.}

\begin{lemma}
\label{lm:02}
{An ideal beam pattern {having identical non-zero
reference gain}
\begin{align*}
 G_{q,p}^{\textrm{ideal}}(\psi_h,\psi_v)&=\frac{Q \Lambda}{M}    \mathbbm{1}_{\cB_{q,p}}(\psi_h,\psi_v),
\end{align*}
{where $\Lambda \doteq \Lambda_h \Lambda_v$ and $\Lambda_a \doteq \pi/\psi^{B}_a$ for $a\in \{ h,v\}$,} satisfies the upper bound of the expected data rate
\begin{align*}
\mathrm{R}^{\mathrm{up}} =  \log_{2} \bigg({1+ \frac{\rho \|\bh \|_2^2  Q  \Lambda }{M } }\bigg).
\end{align*}}
\end{lemma}
\begin{IEEEproof}
{Assuming uniform distribution of  dominant beam directions, the data rate is averaged with respect to the uniformly distributed  beam directions $(\psi_{h},\psi_{v}) \in \cB_{q,p}$. For a conditioned on $\|\bh \|_2^2$, the expected data rate is  bounded as
\begin{align*}
\mathrm{R} &=  \mathrm{E}\Big[  \log_{2} \big({1+ \rho \|\bh \|_2^2 t(\psi_h,\psi_v) }\big) ~\big|~ \| \bh\|_2^2\Big]
\\
&=\frac{\int \int_{\mathcal{B}_{q,p}} \log_{2} \big({1+ \rho \|\bh \|_2^2  t(\psi_h,\psi_v)   }\big)d\psi_h d\psi_v }{\int \int_{\mathcal{B}_{q,p}}d\psi_h d\psi_v  }
\\
&\stackrel{(a)} \leq  \log_{2}\bigg({1+  \frac{  \rho \|\bh \|_2^2 \int \int_{\mathcal{B}_{q,p}}  t(\psi_h,\psi_v)   d\psi_h d\psi_v}{\int \int_{\mathcal{B}_{q,p}}d\psi_h d\psi_v} }\bigg)
\\
&\stackrel{(b)} = \log_{2}\bigg({1+    \frac{ \rho \|\bh \|_2^2 (  2\pi)^2}{M\Delta_h \Delta_v} }\bigg)
\end{align*}
where  $\Delta_a  \doteq   2\psi^{\textrm{B}}_a/{Q_a}$ is defined in (\ref{eq:ebw}). Note that $(a)$ is derived based on Jensen's inequality and $(b)$ is derived based on the constraint in  (\ref{eq:con}). The equality in $(a)$ holds if the $t(\psi_{h},\psi_{v})$ is uniform over $\cB_{q,p}$ {according to}
\begin{align*}
t(\psi_{h},\psi_{v} )&=\frac{(2\pi)^2}{M \Delta_h \Delta_v}=\frac{Q \Lambda }{M}.
\end{align*}
The reference gain is finalized as\footnote{{Any reference gain is upper bounded as $G(\psi_h,\psi_v,\bc)  \leq \big \| \bd_{M_h}(\psi_h) \otimes \bd_{M_v}(\psi_v) \big\|_{2}^{2}\big\| \bc \big\|_{2}^{2}=1$.}} $t(\psi_{h},\psi_{v} ) \doteq \min \big\{1, \frac{Q \Lambda}{M}\big\}$. In this paper, we have $ t(\psi_{h},\psi_{v} ) \leq 1$ because of the assumption $M \geq Q \Lambda$. Finally, the ideal beam pattern  becomes
\begin{align*}
 G_{q,p}^{\textrm{ideal}}(\psi_h,\psi_v)&=\frac{Q \Lambda}{M }  \mathbbm{1}_{\cB_{q,p}}(\psi_h,\psi_v)
\end{align*}
with $(\psi_h,\psi_v) \in \cB_s$ and the corresponding upper bound of the expected data rate is given by}
\begin{align}
\label{eq:up}
\mathrm{R}^{\mathrm{up}} =  \log_{2} \bigg({1+ \frac{\rho \|\bh \|_2^2  Q \Lambda}{M } }\bigg) .
\end{align}
\end{IEEEproof}

\begin{figure}[!t]
\centering
{\includegraphics[width=0.475\textwidth]{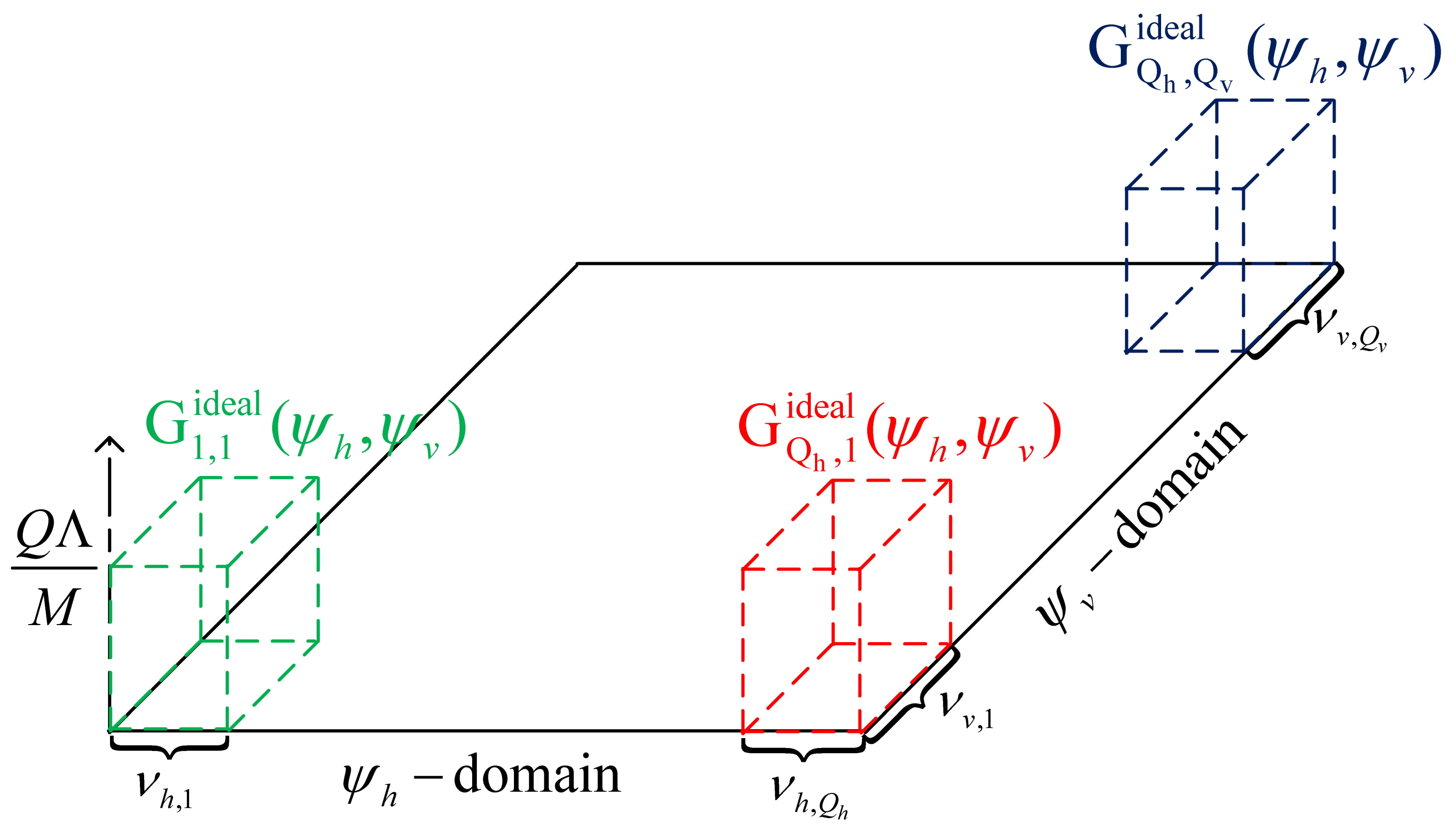}}
\caption{Ideal beam pattern for each beamformer.}
\label{fig:ideal_bp}
\end{figure}

{{\textit{Example ideal beam patterns} are depicted} in Fig. \ref{fig:ideal_bp}. Note that the \textit{ideal beam patterns} will guide the development of the proposed codebook design algorithm.}

\section{Proposed codebook design algorithm}

\subsection{{Problem formulation for beam pattern design}}
In this section, we propose a codebook design algorithm utilizing the  predefined  ideal beam patterns. {The ideal beam patterns are usually unachievable in practice, and we design {hybrid beamformers} that generate beam patterns close to the ideal beam patterns. Beamformers satisfying the hybrid beamforming setup are formed by combination of an analog beamsteering matrix and a baseband beamformer such as
\begin{align*}
\bc_{q,p}=\bF_{q,p}\bv_{q,p}.
\end{align*}
We thus focus on constructing a set of an analog beamsteering matrix and a baseband beamformer that  minimizes the MSE between the ideal beam pattern and the {actual beam patten as}
\begin{align}
\label{eq:opt}
&(\bF^\textrm{opt}_{q,p},\bv^\textrm{opt}_{q,p})
\\
\nonumber
&=\argmin_{{\bF}, {\bv}} \int  \int_{\cB_s} \big|   G_{q,p}^{\textrm{ideal}}(\psi_h,\psi_v) - G(\psi_h,\psi_v, {\bF}{\bv})  \big|^2 d\psi_h d\psi_v,
\end{align}
where the combination of ${\bF}=[\bff_1,\cdots,\bff_N]$ and ${\bv} $ are subject to $\| {\bF}{\bv} \|_2^2=1$.}

{Although we have $Q$ beamformers to optimize, the following lemma shows that {it is possible to generate} other beamformers from one optimized beamformer. {The lemma exploits the phase shifting function defined as
\begin{align}
\label{eq:ps}
\mathrm{T}\big(\bF , \upsilon,\kappa \big) & \doteq \bF  \odot \big(  \tilde{\bd}_M( \upsilon,\kappa )\mathbf{1}_{1,N}\big)
\\
\nonumber
&= \big[\big(\bff_1 \odot   \tilde{\bd}_M( \upsilon,\kappa )\big),\cdots,\big(\bff_N \odot \tilde{\bd}_M(\upsilon,\kappa)\big)\big]
\end{align}
where the normalized array response vector is given by
\begin{align*}
\tilde{\bd}_M( \upsilon,\kappa ) \doteq \frac{\bd_M( \upsilon,\kappa )}{\| \bd_M( \upsilon,\kappa ) \odot \bd_M( \upsilon,\kappa ) \|_2}.
\end{align*}
\begin{lemma}
\label{lm:03}
The beamsteering matrix for the {MSE problem in (\ref{eq:opt})} is obtained by shifting phase directions of  equal gain vectors in the $(1,1)$-th beamsteering matrix such as
\begin{align*}
 \bF^\textrm{opt}_{q,p}= \mathrm{T} \big(\bF^\textrm{opt}_{1,1},  \Delta_h^q , \Delta_v^p \big).
\end{align*}
{The baseband beamformers are all}
\begin{align*}
\bv^\textrm{opt}_{q,p}=\bv^\textrm{opt}_{1,1}.
\end{align*}
{Please see Appendix \ref{sec:A} for the proof.}
\end{lemma}}

{For the rest of this section, we focus on optimizing the $(1,1)$-th beamformer based on Lemma \ref{lm:03}. {It is not practical to consider continuous beam directions for the problem. Therefore, the $b$-th range  $\nu_{a,b}$ in (\ref{eq:ebw}) is quantized with $L_a$ beam directions\footnote{{In the {beam region} $\cB_s$, we consider $L\doteq L_h L_v$ beam directions satisfying $L Q \geq M$.}} {according to}}
\begin{align}
\label{eq:bd}
\psi_a^{b}[\ell] \doteq -{\psi}^{\textrm{B}}_a+ \Delta_a^b + \Delta_a\frac{ \ell-0.5 }{L_a},
\end{align}
where $\ell \in \{ 1, \cdots, L_a\}$, $b \in \{ 1, \cdots, Q_a\}$, and $a \in \{ h,v\}$. Both actual and ideal beam patterns are then represented in vector forms. The quantized beam pattern is
\begin{align*}
&\bG( {\bF} {\bv})\doteq \big[G(\psi^{1}_h[1],\psi^{1}_v[1],  {\bF} {\bv})\cdots G(\psi^{Q_h}_h[L_h],\psi^{Q_v}_v[L_v],  {\bF} {\bv})\big]^{T},
\end{align*}
and the quantized ideal beam pattern is  then
\begin{align}
\label{eq:06}
&\bG^{\textrm{ideal}}_{1,1} \doteq \frac{Q_h \Lambda_h}{M_h} \big(\bee_{h} \otimes \mathbf{1}_{L_{h},1}\big) \otimes  \frac{Q_v \Lambda_v}{M_v} \big(\bee_{v} \otimes \mathbf{1}_{L_{v},1}\big)
\end{align}
where $\bee_{a}$ is the first column vector of  $\bI_{Q_a}$.  The optimization problem is then given by}
\begin{align}
\label{eq:05}
&(\bF_{1,1},\bv_{1,1})=\argmin_{{\bF}, {\bv}} \big\| \bG^{\textrm{ideal}}_{1,1} - \bG( {\bF} {\bv}) \big\|_{2}^{2}.
\end{align}

To get insights on the structure of the  beam patterns {in the optimization (\ref{eq:05}),} we decompose each entry into an arbitrary complex number and its complex conjugate. The ideal beam pattern vector {is  decomposed as}
\begin{align}
\nonumber
\bG^{\textrm{ideal}}_{1,1}\stackrel{(a)}=& \frac{Q_h\Lambda_h}{M_h} \big(\bee_{h} \otimes ( {\bg_h} \odot  {\bg_h^*})\big) \otimes  \frac{Q_v\Lambda_v}{M_v} \big(\bee_{v} \otimes ( {\bg_v} \odot  {\bg_v^*})\big)
\\
\nonumber
=& \big( \sqrt{{Q\Lambda}/{M}} (\bee_{h} \otimes  {\bg_h}) \otimes (\bee_{v} \otimes  {\bg_v}) \big)
\\
\label{eq:07}
&~~~~~~~\odot  \big(\sqrt{{Q\Lambda}/{M}}  (\bee_{h}\otimes  {\bg_h}) \otimes (\bee_{v} \otimes  {\bg_v}) \big)^*.
\end{align}
{Note that $(a)$ is derived because the all ones vector  in (\ref{eq:06}) can be decomposed into any equal gain vector and its element-wise complex conjugate as $\mathbf{1}_{L_a,1}= {\bg_a} \odot  {\bg_a^*}$, where $\bg_a$ has unit gain entries.} Similarly, the actual beam pattern is decomposed as
\begin{align}
\label{eq:act}
\bG( {\bF} {\bv})&=\big( \big(\bD_h^H \otimes \bD_v^H \big) {\bF} {\bv} \big) \odot \big( \big(\bD_h^H \otimes \bD_v^H \big)  {\bF} {\bv} \big)^{*}
\end{align}
with the set of array vectors for quantized directions in (\ref{eq:bd})
\begin{align*}
\bD_{a}&=\big [\bD_{a,1}, \cdots, \bD_{a,Q_a} \big] \in \mathbb{C}^{M_a \times L_aQ_a},
\\
\bD_{a,b} &= [\bd_{M_a}(\psi_a^{b}[1]), \cdots, \bd_{M_a}(\psi_a^{b}[L_a])] \in \mathbb{C}^{M_a \times L_a}.
\end{align*}

{Unfortunately, there exists no {closed-form solution} for the minimization problem in (\ref{eq:05}) because the entries in $\bG( {\bF} {\bv})$ and $\bG^{\textrm{ideal}}_{1,1}$ are found in the form of absolute square of a complex number. To access a feasible, but usually suboptimal, solution, we reformulate our problem that compares the decomposed vectors in (\ref{eq:07}) and (\ref{eq:act}) for a given $(\bg_h,\bg_v)$
\begin{align*}
&\big(\bF_{|{\bg_h,\bg_v}}, \bv_{|{\bg_h,\bg_v}}\big)=
\\
&\argmin_{ {\bF}, {\bv}} \bigg \| {\beta} \big(\bD_h^H \otimes \bD_v^H \big) {\bF} {\bv}-\sqrt{\frac{Q \Lambda}{M}}(\bee_{h} \otimes  {\bg_h}) \otimes (\bee_{v} \otimes  {\bg_v})  \bigg \|_2^2
\end{align*}
where ${\beta} \in \mathbb{C}$ is a normalization constant.} {We  compute the normalization constant ${\beta}$ by differentiating the {objective function}  over $ {\beta}^{*}$ as
\begin{align*}
&\frac{\partial}{\partial \beta^*}\bigg \| {\beta} \big(\bD_h^H \otimes \bD_v^H \big) {\bF} {\bv}-\sqrt{\frac{Q \Lambda}{M}}(\bee_{h} \otimes  {\bg_h}) \otimes (\bee_{v} \otimes  {\bg_v})  \bigg \|_2^2
\\
&=\beta\big\| (\bD_h^H \otimes \bD_v^H ) \bF \bv \big\|_2^2
\\
&~~~~~~~~~~~~~~~-\sqrt{\frac{Q \Lambda}{M}} (\bF \bv)^H\big(  \bD_h(\bee_{h} \otimes {\bg_h}) \otimes \bD_v(\bee_{v} \otimes {\bg_v})  \big)
\end{align*}
based on Wirtinger derivatives, which simplify differentiation in complex variables \cite{Ref_Rem91}. The complex gain that minimizes the {objective function} is then given by}
\begin{align*}
\hat{\beta}&=\frac{\sqrt{\frac{Q \Lambda}{M}}({\bF}{\bv})^H \big(  \bD_h(\bee_{h} \otimes {\bg_h}) \otimes \bD_v(\bee_{v} \otimes {\bg_v})  \big)  }{\big\| \big( \bD_h^H \otimes \bD_v ^H\big) {\bF}{\bv} \big\|_2^2}
\\
&\stackrel{(a)}=\frac{\sqrt{\frac{Q \Lambda}{M}}({\bF}{\bv})^H \big(\bD_{h,1}{\bg_h} \otimes  \bD_{v,1}{\bg_v}\big) }{\big\| \big( \bD_h^H \otimes \bD_v^H \big) {\bF}{\bv} \big\|_2^2},
\end{align*}
where $(a)$ is derived because $ \bD_a(\bee_{a} \otimes {\bg_a}) = \bD_{a,1}{\bg_a}$. {By plugging $\hat{\beta}$ into the object function, the problem formulation is simplified to}
\begin{align}
\nonumber
\big(\bF_{|{\bg_h,\bg_v}},\bv_{|{\bg_h,\bg_v}}\big)&=\argmax_{{\bF}, {\bv}} \frac{\big|  \big(\bD_{h,1}{\bg_h} \otimes  \bD_{v,1}{\bg_v}\big)^H        {\bF}{\bv}  \big|^2 } {\big\| \big( \bD_h ^H\otimes \bD_v ^H\big) {\bF}{\bv} \big\|_2^2}
\\
\label{eq:10}
&\stackrel{(a)}=\argmax_{{\bF}, {\bv}} \big|      \big(\bD_{h,1}{\bg_h} \otimes  \bD_{v,1}{\bg_v}\big)^H        {\bF}{\bv} \big|^2,
\end{align}
where $(a)$ is derived because $\bD_a\bD_a^H=\frac{L_aQ_a}{M_a}\bI_{M_a}$ and the denominator is fixed to $ {\big\| \big( \bD_h ^H\otimes \bD_v^H \big) {\bF}{\bv} \big\|_2^2}=\frac{LQ}{M}$ for any ${\bF} {\bv}$ satisfying the power constraint $\| {\bF} {\bv} \|_2^2=1$.

{We denote the solution for the problem in (\ref{eq:10})
\begin{align*}
\bc_{|{\bg_h,\bg_v}} = \bF_{|{\bg_h,\bg_v}} \bv_{|{\bg_h,\bg_v}}
\end{align*}
as a beamformer candidate for a given $(\bg_h,\bg_v)$.} We assume that equal gain vectors {in the domain $a \in \{h,v\}$ are} subject to the constrained set\footnote{Any set of equal gain vectors $(\bg_h,\bg_v) \in \cG_{L_h}^I \times \cG_{L_v}^I$ construct $\bG^{\textrm{ideal}}_{1,1}$. The  entries in ${\bg_a}$ have the fixed absolute value, while the phase {can be arbitrary in}  $\cZ_I$, defined in (\ref{eq:phase}).}
\begin{align}
\label{eq:cg}
\cG_{L_a}^{I} = \big\{{\bg} \in \mathbb{C}^{L_a}: ({\bg})_{\ell}=e^{jz},~ z \in \cZ_I \big\}
\end{align}
with $(\bg)_{1}=1$. In the proposed algorithm, we generate beamformer candidates over $(\bg_h,\bg_v) \in \cG_{L_h}^I \times \cG_{L_v}^I$. Each beamformer candidate is a feasible solution  accomplishing the minimization objective in (\ref{eq:opt}). Note that $I^{L_h+L_v-2}$ beamformer candidates are computed because each pair of equal gain vectors, which is the combination of $I^{L_h-1}$ and $I^{L_v-1}$ vectors in both domains,  produces a beamformer.

\subsection{{OMP algorithm constructing beamformer candidates}}
{We now solve the maximization problem in (\ref{eq:10}) to generate each beamformer candidate for a given $(\bg_h,\bg_v)$. An optimal solution to the problem in (\ref{eq:10})  is computed as}
\begin{align*}
\tilde{\bc}_{|{\bg_h,\bg_v}} \doteq \frac{\bD_{h,1}{\bg_h} \otimes  \bD_{v,1}{\bg_v}}{\big \| \bD_{h,1}{\bg_h} \otimes  \bD_{v,1}{\bg_v}\big\|_2}.
\end{align*}
However,  $\bF$ and  $\bv$ may not be able to construct the optimal beamformer because column vectors in $\bF$ are subject to the equal gain subset $\cB_{M}$ in (\ref{eq:equal_set1}).

\begin{algorithm}
  \caption{Beamformer design based on the OMP}
  \label{Al:01}
  \begin{algorithmic}
\State \textbf{Initialization}~For a given $({\bg_h},{\bg_v}) \in \cG_{L_h}^I \times \cG_{L_v}^I$
\State 1:~Optimal beamformer~$\tilde{\bc}_{|{\bg_h,\bg_v}} \doteq \frac{\bD_{h,1}{\bg_h} \otimes  \bD_{v,1}{\bg_v}}{ \| \bD_{h,1}{\bg_h} \otimes  \bD_{v,1}{\bg_v} \|_2}$
\State 2:~Define an initial residual vector~$\br_{0} \doteq \tilde{\bc}_{|{\bg_h,\bg_v}}$
\State 3:~Define an initial empty matrix~${\bF}_{0}$

\State \textbf{Iterative update}
\State 4:~~\textbf{for}~$1 \leq n \leq N$
\State 5:~~{Choose equal gain vector~${\bff}_{n}=\frac{1}{\sqrt{M}}\exp{(j  \angle \br_{n-1})}$}
\State 6:~~{Quantize each phase element~$ \angle({\bff_n})_{m} \in \mathcal{Z}_{2^B}$}
\State 7:~~Update beamsteering matrix~${\bF}_{n}=[{\bF}_{n-1},{\bff}_{n}] \in \mathbb{C}^{M \times n}$
\State 8:~~Update baseband beamformer
\State ~~~~${\bv}_{n}=\frac{\mathfrak{v}_{\textrm{max}} \{(\bF_n^H \bF_n)^{-1}\bF_n^H   (\mathbf{\Gamma}_{h} \otimes \mathbf{\Gamma}_{v})   \bF_n \}}{\big\|\bF_n \mathfrak{v}_{\textrm{max}} \{(\bF_n^H \bF_n)^{-1}\bF_n^H  (\mathbf{\Gamma}_{h} \otimes \mathbf{\Gamma}_{v})    \bF_n \} \big\|_2} \in \mathbb{C}^{n}$,
\State ~~~~where~$\mathbf{\Gamma}_{a} \doteq \bD_{a,1}{\bg_a}{\bg_a^H}\bD_{a,1}^H  \in \mathbb{C}^{M_a \times M_a}$
%\State 8:~~~$\bc_n={\bF}_{n} {\bv}_{n}$
\State 9:~~Update residual vector $\br_{n}=\tilde{\bc}_{|{\bg_h,\bg_v}} - {\bF}_{n} {\bv}_{n}$

\State 10:~\textbf{end~for}

\State \textbf{Final update}
\State 11:~Update analog beamsteering matrix~${\bF}_{|{\bg_h},{\bg_v}}={{\bF}}_{N}$
\State 12:~Update baseband beamformer~${\bv}_{|{\bg_h},{\bg_v}}={{\bv}}_{N}$

\State \textbf{Final output}
\State 13:~Compute codeword candidate~${\bc}_{|{\bg_h},{\bg_v}}={\bF}_{|{\bg_h},{\bg_v}}{\bv}_{|{\bg_h},{\bg_v}}$
  \end{algorithmic}
\end{algorithm}

{We first compute a beamsteering vector $\bff_{n}$ at the $n$-th update. To remove the $(n-1)$-th residual vector\footnote{The initial residual vector is defined as $\br_{0}\doteq \tilde{\bc}_{|{\bg_h,\bg_v}}$.}
\begin{align*}
\br_{n-1}=\tilde{\bc}_{|{\bg_h,\bg_v}} - {\bF}_{n-1} {\bv}_{n-1}
\end{align*}
that is not suppressed in the previous  update, the beamsteering vector is computed as  \cite{Ref_Lov03}, ${\bff}_{n}=\frac{1}{\sqrt{M}}\exp{(j \bm{\theta}_n)}$, where
\begin{align*}
\bm{\theta}_n &\in \argmax_{\bm{\vartheta} \in [0,2\pi)^M} \big|\br_{n-1}^H e^{j \bm{\vartheta} } \big|^2.
\end{align*}
The optimal phase vector $\bm{\theta}_n$ does not have  a single unique solution because
\begin{align*}
\big|\br_{n-1}^H \exp{(j \bm{\theta}_n)}\big|^2=\big|\br_{n-1}^H \exp{(j \bm{\theta}_n)} \exp(j \xi )\big|^2
\end{align*}
for any phase angle $\xi \in [0,2\pi)$. Although the optimal phase vector is given by $\bm{\theta}_n =   \angle \br_{n-1} + \xi$ in \cite{Ref_Lov03}, we define $\bm{\theta}_n \doteq \angle \br_{n-1}$ with $\xi=0$ for simplicity. Note that $\angle  \br_{n-1}  \in [0,2\pi)^M$ is the function that returns each element phase of $\br_{n-1} \in \mathbb{C}^M$ in a vector form.} {Assuming the digitally-controlled RF phase shifter \cite{Ref_Hur13},} each element phase in $\bff_{n} $ is quantized with {the set of quantized phases $\mathcal{Z}_{2^B}$ in (\ref{eq:phase}).} The  beamsteering matrix at the $n$-th update is then given by ${\bF}_{n}=[{\bF}_{n-1},{\bff}_{n}]  \in \mathbb{C}^{M \times n}$.}

\begin{figure*}[!t]
\normalsize
\centering
\subfigure[Proposed codebook]{\includegraphics[width=0.3285\textwidth]{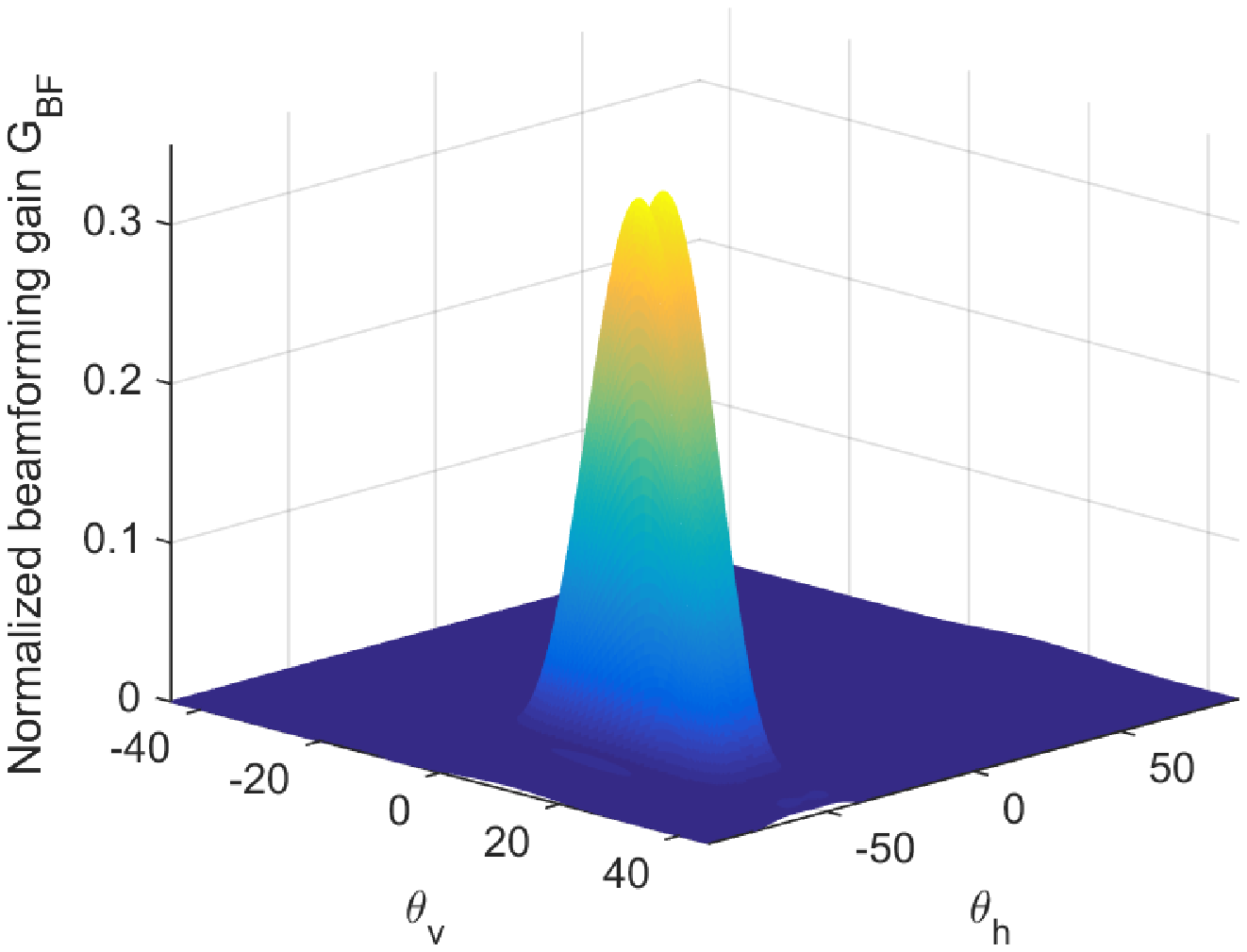}}
\hfil
\subfigure[Codebook in \cite{Ref_Hur13}]{\includegraphics[width=0.3285\textwidth]{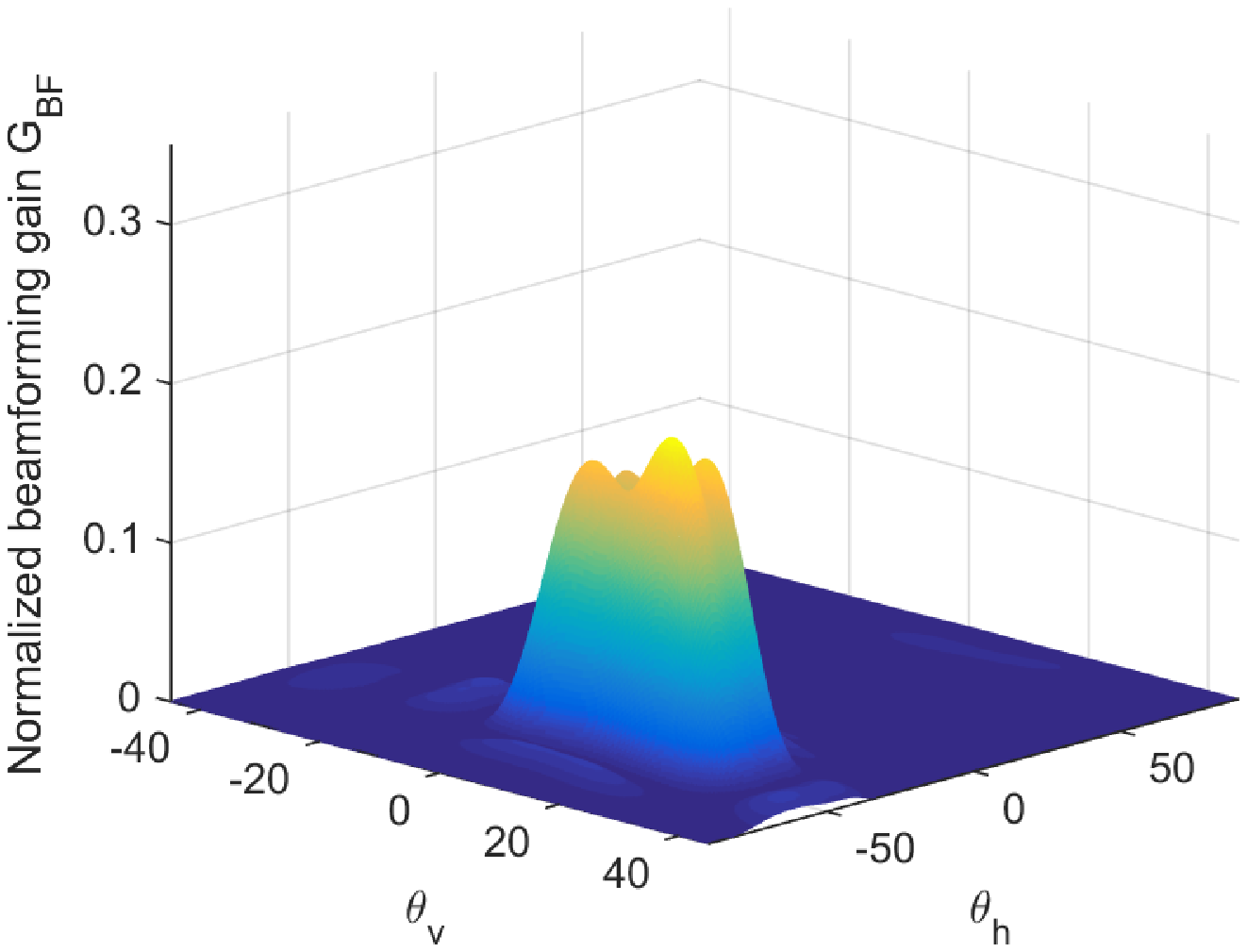}}
\hfil
\subfigure[Codebook with $\bar{\bg}_h~\&~\bar{\bg}_v$ \cite{Ref_Alk14}]{\includegraphics[width=0.3285\textwidth]{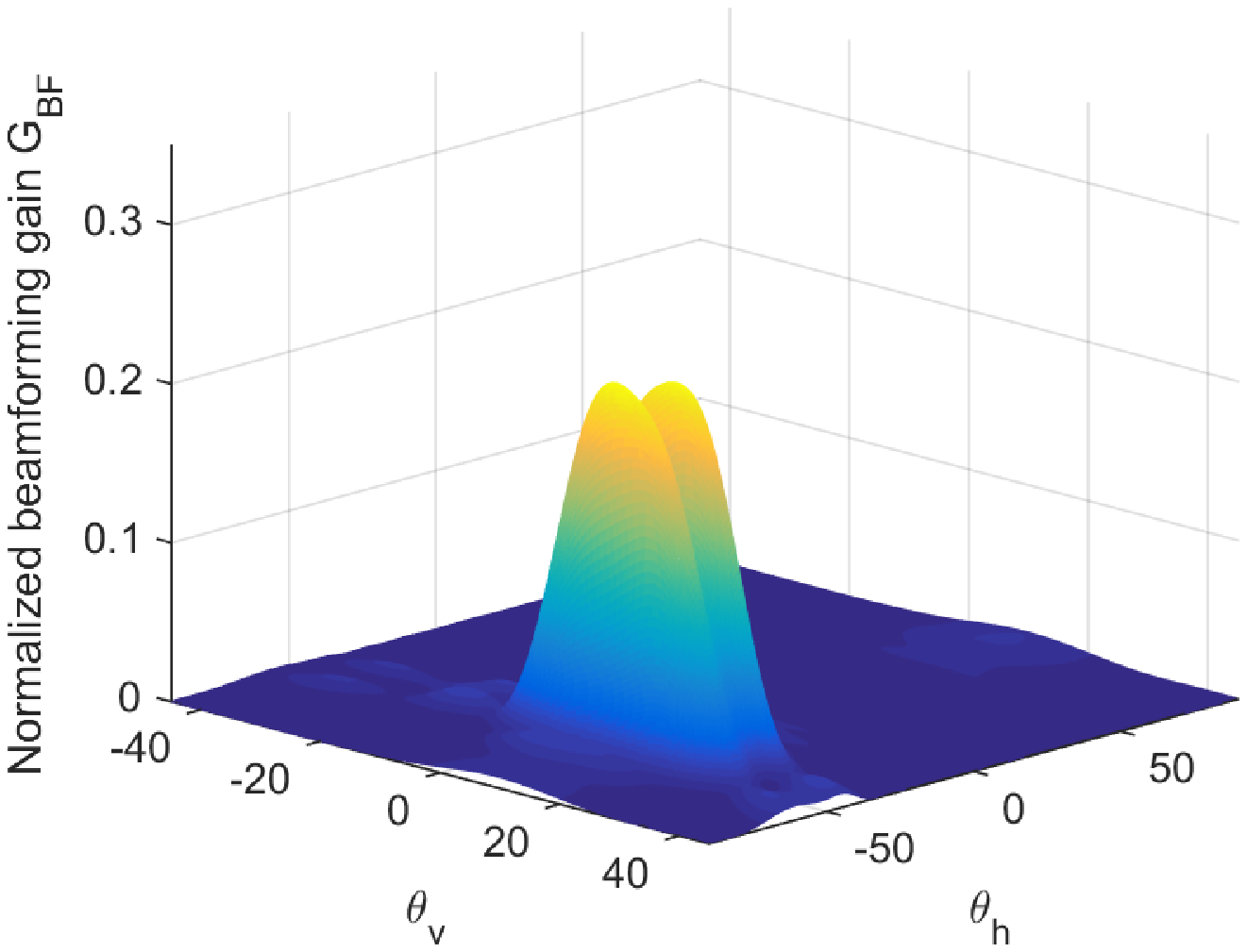}}
\caption{Normalized beamforming gains $\mathrm{G}_{\mathrm{BF}}(\theta_h,\theta_v,\bc_{2,2})$ with $(M_h,M_v)=(16,8),~(Q_h,Q_v)=(8,4),~N=4$.}
\label{fig:bp_one}
\end{figure*}

{Similar to the beamformer designs in \cite{Ref_Alk14,Ref_Aya14}, the problem in (\ref{eq:10}) can be solved by} utilizing the OMP algorithm in \cite{Ref_Tro07,Ref_Reb02}. Based on the OMP algorithm, we choose each equal gain vector  one-by-one and update the baseband beamformer iteratively in such a way that a combination of ${\bF}_{|{\bg_h,\bg_v}}$ and ${\bv}_{|{\bg_h,\bg_v}}$  minimizes  the norm of residual vector}
\begin{align*}
\br \doteq \tilde{\bc}_{|{\bg_h,\bg_v}}-\bF_{|{\bg_h,\bg_v}} \bv_{|{\bg_h,\bg_v}}.
\end{align*}

{A baseband beamformer $\bv_{n}$ is then computed by solving the maximization problem (\ref{eq:10}) over $\bv \in \mathbb{C}^{n}$.} To compute the beamformer that satisfies the power constraint of hybrid beamforming system  $\|{\bF}{\bv}\|_{2}^{2}=1$, the maximization problem is rewritten by changing dummy variables as $\bv=\frac{\bu}{\| \bF_n\bu \|_2}$ for ${\bu} \in \mathbb{C}^n$. {For a given $(\bg_h,\bg_v)$ and $\bF_{n}$, the baseband beamformer at the $n$-th update is computed based on the generalized Rayleigh quotient solution in \cite{Ref_Bor98}, such as $\bv_{n}=\frac{\bu_{n}}{\| \bF_{n}\bu_{n} \|_2}$, where}
\begin{align}
\nonumber
\bu_{n}&=\argmax_{{\bu} \in  \mathbb{C}^{n}} \frac{{\bu}^H \big(\bF_n^H   \big( \mathbf{\Gamma}_{h} \otimes \mathbf{\Gamma}_{v} \big) \bF_{n}\big){\bu}}{ {\bu}^H (\bF_n^H \bF_n) {\bu} }
\\
\label{eq:11}
&=\mathfrak{v}_{\textrm{max}}\big\{(\bF_n^H \bF_n)^{-1}\bF_n^H  \big( \mathbf{\Gamma}_{h} \otimes \mathbf{\Gamma}_{v} \big)  \bF_n \big\}
\end{align}
with $\mathbf{\Gamma}_{a} \doteq \bD_{a,1}{\bg_a}\bg_a^H\bD_{a,1}^H \in \mathbb{C}^{M_a \times M_a}$.

{The  beamformer at the $n$-th update becomes $\bc_n={\bF}_{n} {\bv}_{n}$ and the residual vector $\br_{n}=\tilde{\bc}_{|{\bg_h,\bg_v}} - {\bF}_{n} {\bv}_{n}$ is updated for the following update steps.} The iterative process is summarized in Algorithm~\ref{Al:01}. {Each beamformer candidate is formed by combination of the updated solution, i.e.,  ${\bF}_{|{\bg_h},{\bg_v}}$ and $\bv_{|{\bg_h,\bg_v}}$,
\begin{align*}
\bc_{|{\bg_h,\bg_v}}=\bF_{|{\bg_h,\bg_v}}\bv_{|{\bg_h,\bg_v}}
\end{align*}
for a given  $(\bg_h,\bg_v) \in \cG^{I}_{L_h} \times \cG^{I}_{L_v}$.}

\subsection{{Final beamformer selection with MSE minimization}}

{The beamformer that generates the beam pattern close to the $(1,1)$-ideal beam pattern is chosen as}
\begin{align*}
\bc_{1,1}={\bF}_{| \hat{\bg}_h,\hat{\bg}_v}{\bv}_{| \hat{\bg}_h,\hat{\bg}_v}
\end{align*}
where a set of the beamsteering matrix and the baseband beamformer that accomplishes the minimization objective in (\ref{eq:opt}) is given by\footnote{{Because the codebook is constructed offline, the brute force search in (\ref{eq:likelihood}) is performed only once and has no impact on system operation.}}
\begin{align}
\label{eq:likelihood}
&(\hat{\bg}_h,\hat{\bg}_v)=\argmin_{\bg_h , \bg_v} \big\| \bG^{\textrm{ideal}}_{1,1} - \bG( {\bF}_{|\bg_h,\bg_v} {\bv}_{|\bg_h,\bg_v}) \big\|_{2}^{2}
\end{align}
over $(\bg_h , \bg_v) \in \cG^{I}_{L_h} \times \cG^{I}_{L_v}$. {Increasing $I$ and $L_a$ ensures a large number of beamformer candidate for the optimization problem in (\ref{eq:opt}), while it imposes a heavy computational complexity. To construct $\cG_{L_a}^{I}$ in (\ref{eq:cg}), we thus consider limited numbers of $I$ and $L_a$.}

\begin{figure*}[!t]
\normalsize
\centering
\subfigure[Proposed codebook {(Average reference gain $0.553$)}]{\label{fig:bp_more_a}\includegraphics[width=0.3525\textwidth]{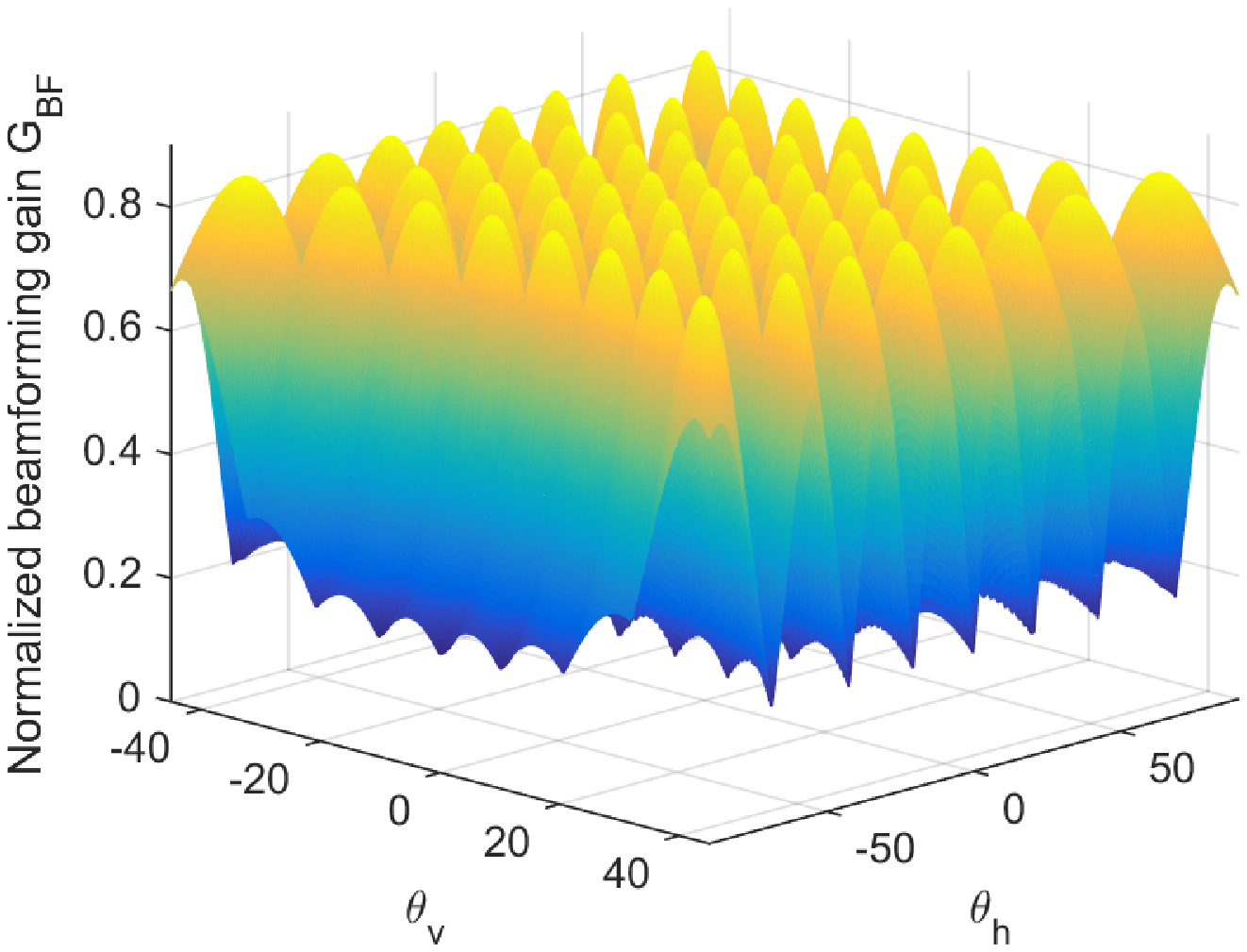}}
\hfil
\subfigure[Proposed codebook with $\gamma=7.5\%$ {(Average reference gain $0.467$)}]{\label{fig:bp_more_b}\includegraphics[width=0.3525\textwidth]{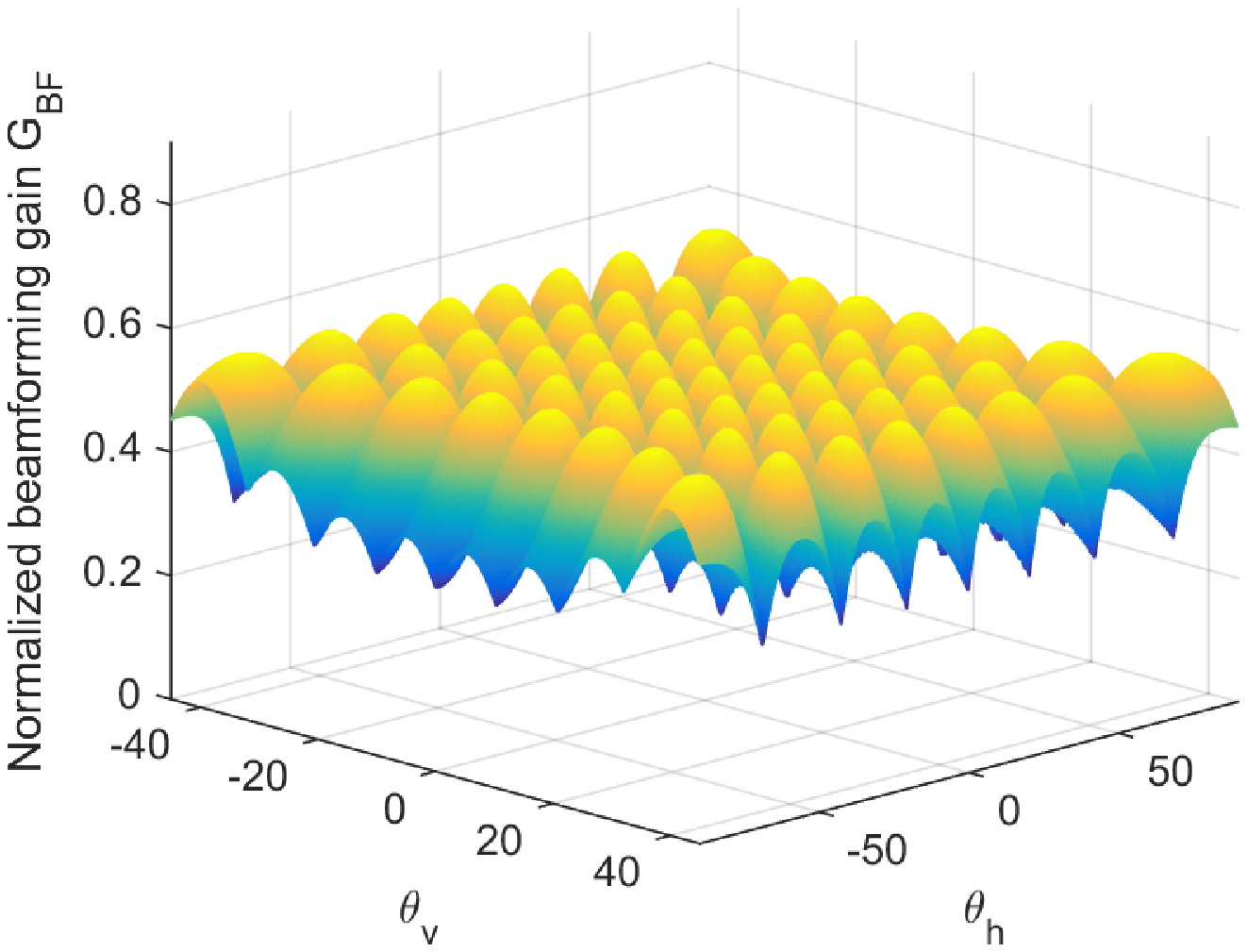}}
\hfil
\subfigure[Codebook in \cite{Ref_Hur13} {(Average reference gain $0.397$)}]{\label{fig:bp_more_c}\includegraphics[width=0.3525\textwidth]{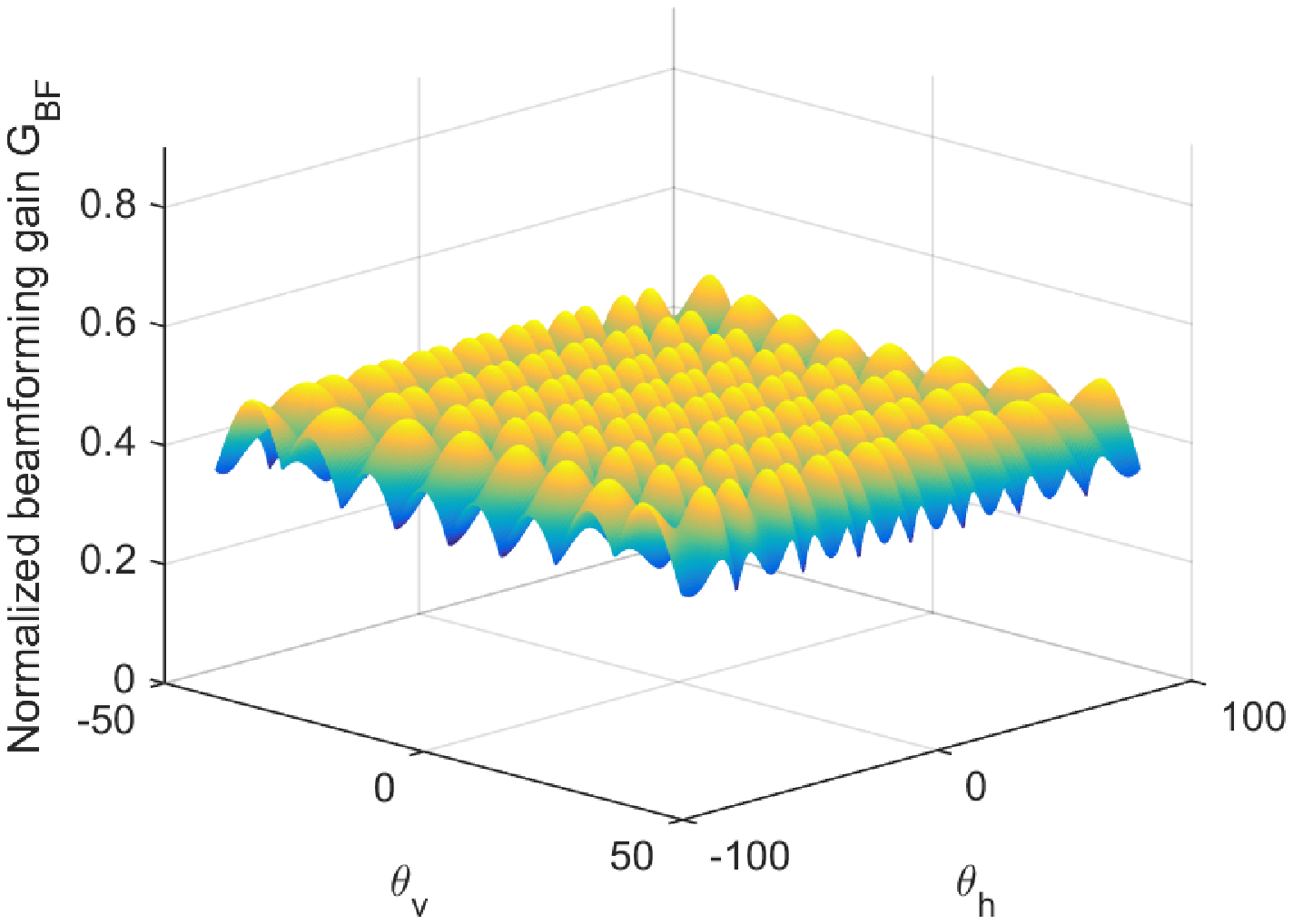}}
\hfil
\subfigure[Codebook with $\bar{\bg}_h~\&~\bar{\bg}_v$ \cite{Ref_Alk14} {(Average reference gain $0.440$)}]{\label{fig:bp_more_d}\includegraphics[width=0.3525\textwidth]{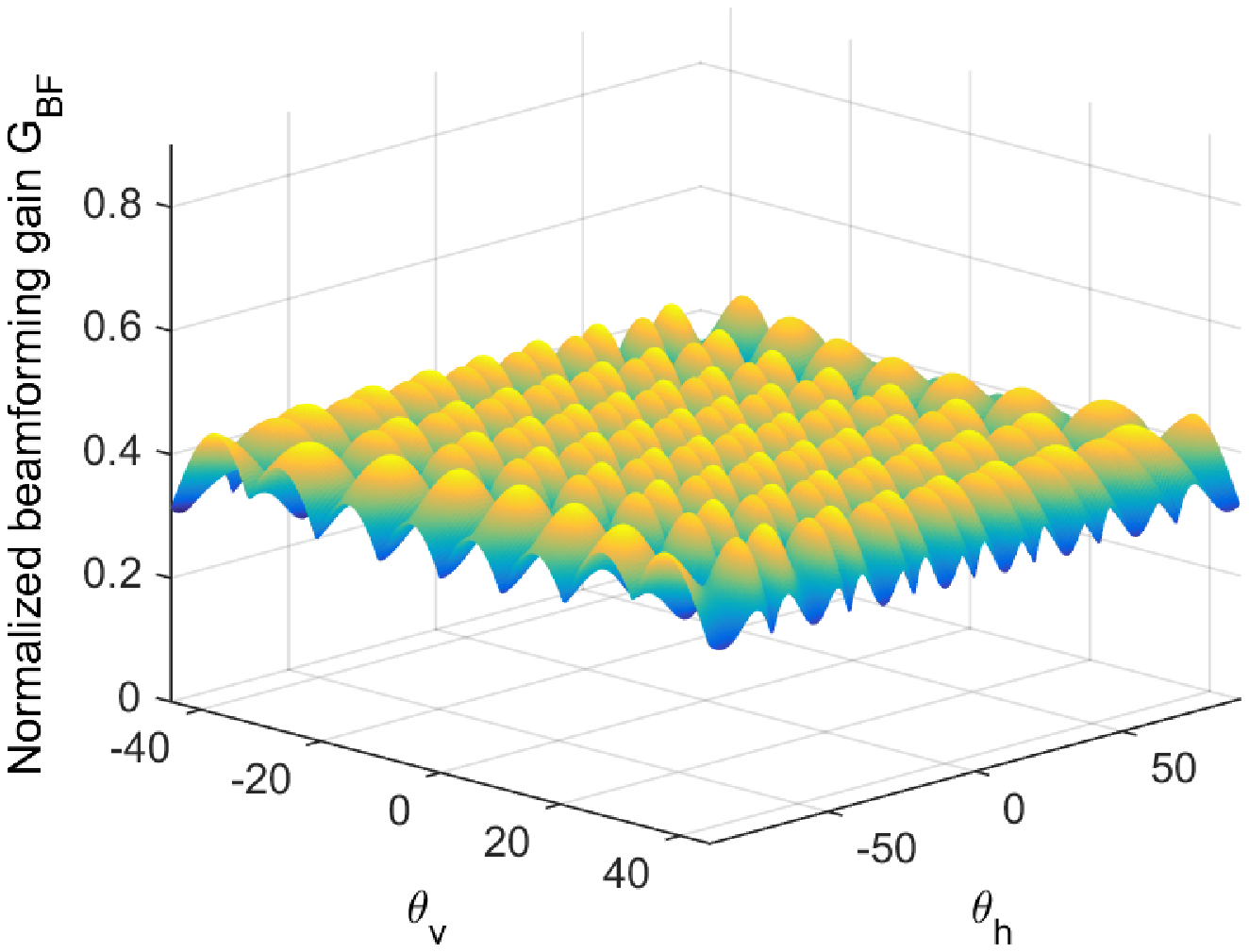}}
\caption{Codebook examples with $(M_h,M_v)=(12,6),~(Q_h,Q_v)=(8,8),~N=4$.}
\label{fig:bp_more}
\end{figure*}

{Finally, all other beamformers in the proposed  codebook $\mathcal{C}=\big\{\bc_{1,1}\cdots\bc_{Q_h,Q_v} \big\}$ can be derived based on Lemma \ref{lm:03} such as}
\begin{align}
\label{eq:12}
\bc_{q,p}=  \mathrm{T} \big( \bF_{| \hat{\bg}_h,\hat{\bg}_v}, \Delta_h^q,\Delta_v^p \big) \bv_{| \hat{\bg}_h,\hat{\bg}_v}.
\end{align}

\subsection{{Overlapped {beam regions}}}
\label{sec:overlap}
{In most beam alignment approaches, the beamformer generating the largest beamforming gain  for data transmission is chosen among the consecutive beamformers in the overlapped {beam region}.} Practical beamformers, which are designed based on  non-overlapping {{beam regions}  in (\ref{eq:BS}),} generate beam patterns having minimal overlap between neighboring beams. This might not be desirable when tying to cover a geographic region with adequate quality of service. {To alleviate sharp dips between consecutive beams, we thus widen} the {beam region} of each beamformer by allowing a guard band  $\tilde{\cB}_{q,p} \doteq \tilde{\nu}_{q}^{h} \times \tilde{\nu}_{p}^{v}$, where
\begin{align*}
\tilde{\nu}_{b}^{a}& \doteq   -{\psi}^{\textrm{B}}_a+ \Delta_a^b+ \Delta_a\big[-\gamma      , 1 + \gamma   \big),
\end{align*}
$\Delta_a$ is the non-overlapped beam-width of each beamformer in (\ref{eq:ebw}), and $\gamma \Delta_a $ is the overlapped beam-width of the guard band, which is defined by the design parameter $\gamma$. Because each beamformer covers the widened {beam region} {including the guard band,} an ideal beam pattern may have a low  non-zero reference gain. Thus, the ideal beam pattern is redefined as %depicted in Fig. \ref{fig:refine_bp} where
\begin{align*}
 \tilde{G}_{q,p}^{\textrm{ideal}}(\psi_h,\psi_v)&=\frac{Q \Lambda}{M(1+2\gamma)^2}\mathbbm{1}_{\tilde{\cB}_{q,p}}(\psi_h,\psi_v).
\end{align*}
Beamformers alleviating the sharp dips are also computed based on the proposed codebook design algorithm by plugging the redefined beam pattern into the optimization problem.

\section{Simulation Results}
\label{sec:SR}
{In this section, numerical results are presented to verify the data rate performances of proposed codebook design algorithm.} In this paper, four RF chains and a six bit phase control register, i.e., $N=4$, $B=6$, are considered for hybrid beamforming architectures. The beamforming codebook $\mathcal{C}=\{\bc_{1,1}\cdots\bc_{Q_h,Q_v} \}$ consisting of $Q=Q_hQ_v$ beamformers is designed as in (\ref{eq:12}). %based on the OMP algorithm summarized in Algorithm \ref{Al:01}.
For the minimization problem in (\ref{eq:likelihood}), the equal gain sets $\cG_{L_h}^{I}$ and $\cG_{L_v}^{I}$ are defined with parameters $L_h=8$, $L_v=8$, and $I=3$. In addition, we consider $20$ directions in each beam-width $\nu_{a,b}$ to compute the MSE between the beamformer candidate's beam pattern and the ideal beam pattern. %In addition, we consider $30$ beam directions in each beam-width $\nu_b^a$ to compute the MSE between the beamformer candidate's beam pattern and the ideal beam pattern in (\ref{eq:likelihood}).

\subsection{{Beam patterns of codebook examples}}
In Figs. \ref{fig:bp_one} and \ref{fig:bp_more}, we compare the beam patterns of the proposed codebook and the codebooks in \cite{Ref_Hur13,Ref_Alk14}, {and the 2D KP codebook in \cite{Ref_3GPP150560}}. {The ULA codebook in \cite{Ref_Hur13} is extended to a 2D UPA codebook  by maximizing a minimum reference gain in each target {beam region} $\cB_{q,p}$.} The codebook in \cite{Ref_Alk14} is extended to UPA structures with a single set of equal gain vectors
\begin{align}
\label{eq:single_set}
(\bar{\bg}_h,\bar{\bg}_v) \doteq \Big(\mathbf{1}_{ \lceil{256}/{Q_h},1  \rceil},\mathbf{1}_{  \lceil {256}/{Q_v},1  \rceil}\Big).
\end{align}
{Finally, the 2D KP codebook is given by $\cD_{h} \times \cD_{v}$  where the  discrete Fourier transform (DFT) codebook in the domain $a \in \{ h,v\}$ is defined as}
\begin{align*}
\cD_{a} \doteq \big\{ \bd_{M_a}({2\pi}/{Q_a}),\cdots, \bd_{M_a}({2\pi Q_a}/{Q_a})\big\}.
\end{align*}

\begin{figure*}[!t]
\normalsize
\centering
\subfigure[$(M_h,M_v)=(12,6),~(Q_h,Q_v)=(8,4)$]{\includegraphics[width=0.4475\textwidth]{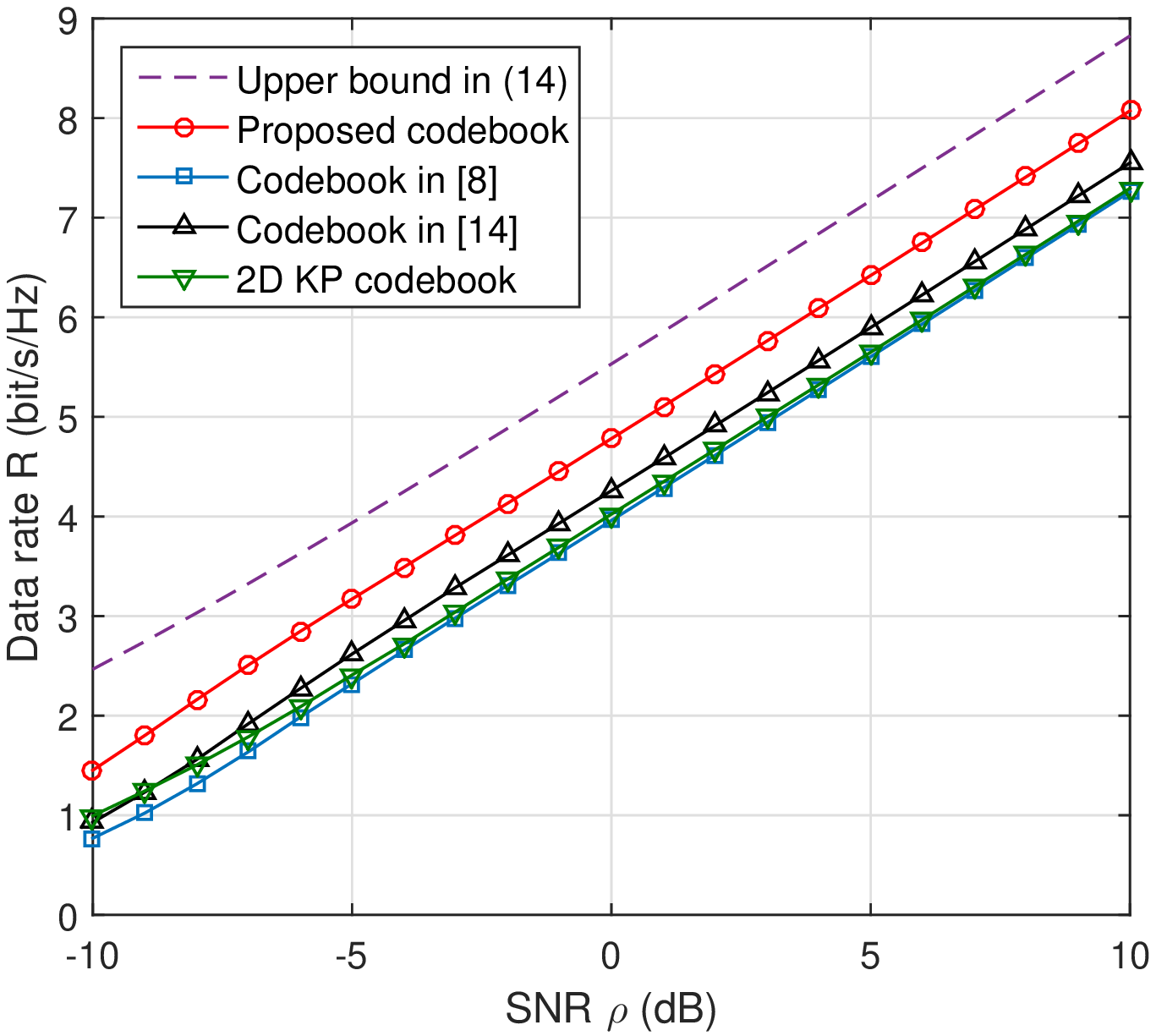}}
\hfil
\subfigure[$(M_h,M_v)=(16,8),~(Q_h,Q_v)=(8,4)$]{\includegraphics[width=0.4475\textwidth]{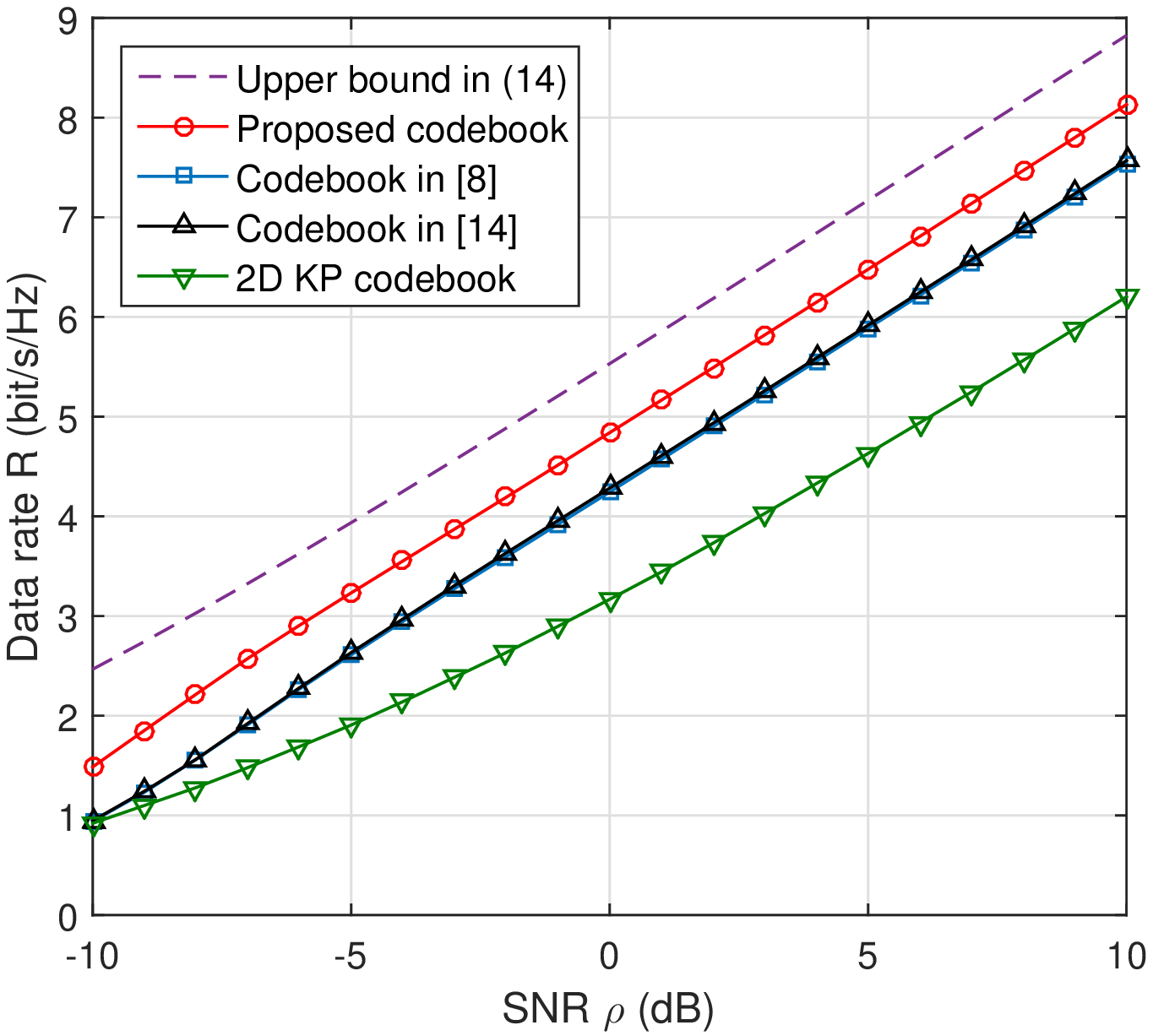}}
\hfil
\subfigure[$(M_h,M_v)=(20,10),~(Q_h,Q_v)=(8,4)$]{\includegraphics[width=0.4475\textwidth]{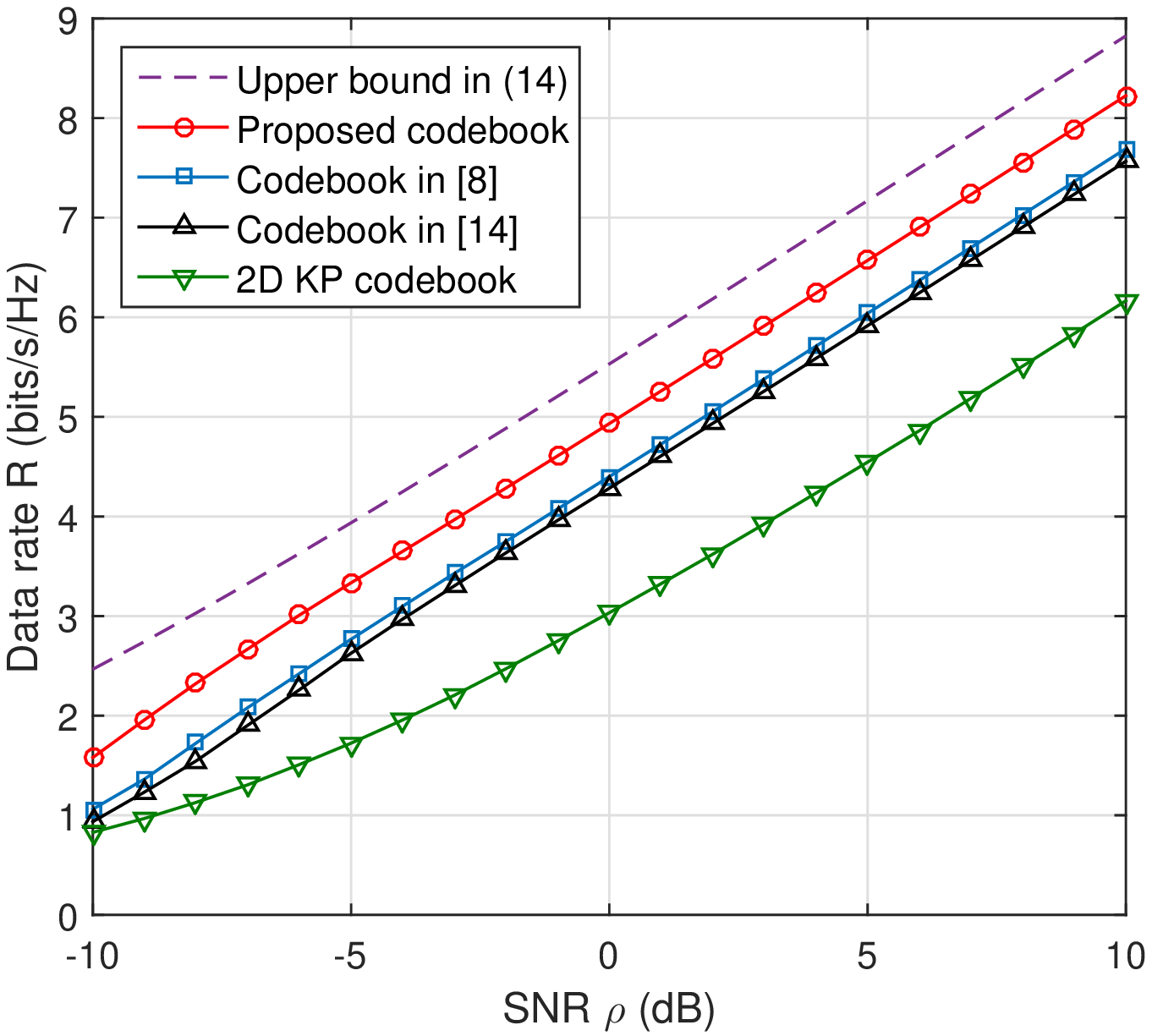}}
\hfil
\subfigure[$(M_h,M_v)=(24,12),~(Q_h,Q_v)=(8,4)$]{\includegraphics[width=0.4475\textwidth]{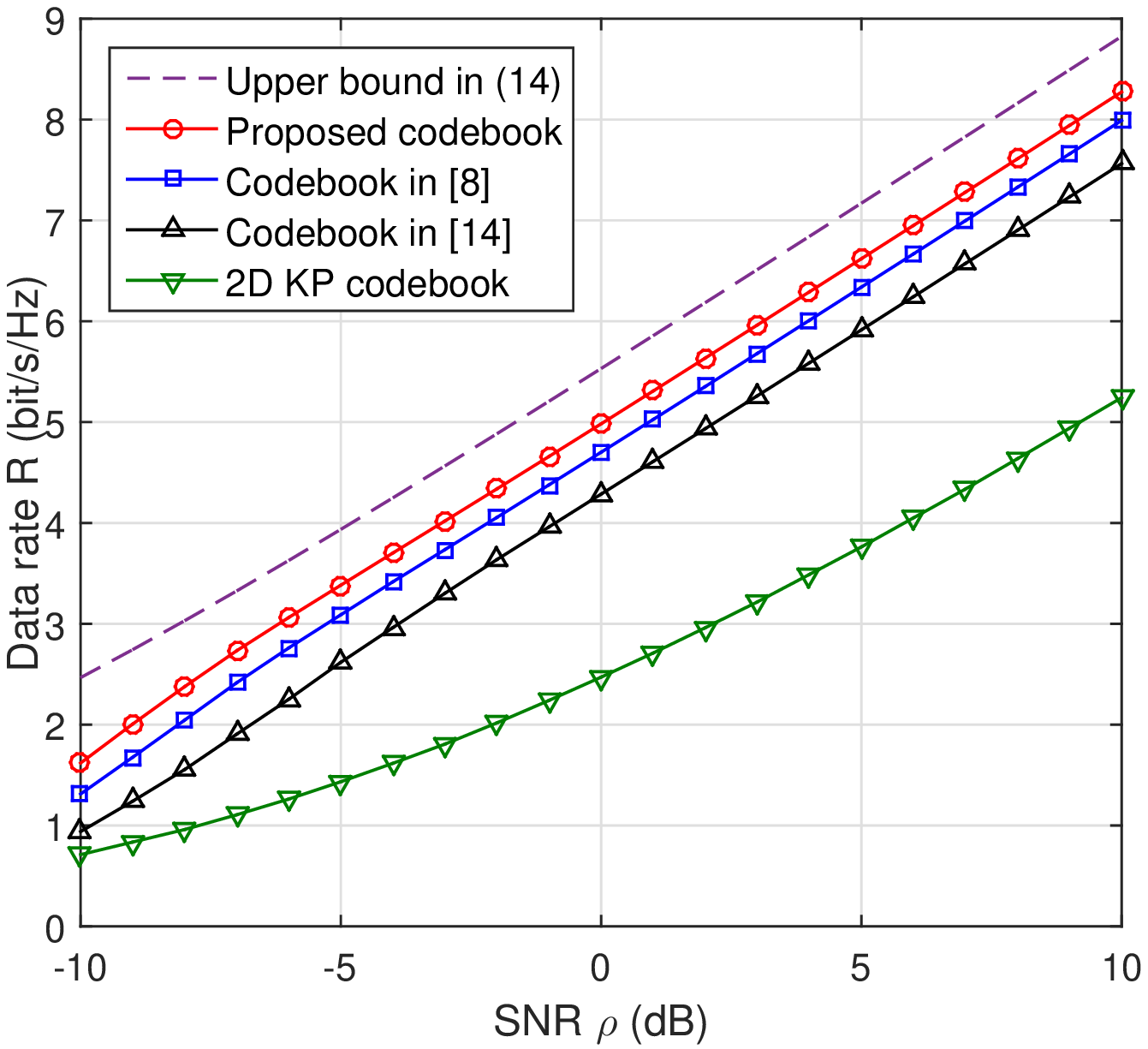}}
\caption{{Data rate based on the channel vectors formed by LOS and NLOS paths.}}
\label{fig:bg_single}
\end{figure*}

{In Fig. \ref{fig:bp_one},} we compare the beam patterns of a single beamformer by using the reference gain defined as
\begin{align*}
  &\mathrm{G}_{\mathrm{BF}}(\theta_h,\theta_v,\bc_{q,p})= \big|  \bc_{q,p}^H \bd_M\big( \pi \sin \theta_h \cos \theta_v, \pi \sin \theta_v \big) \big|^2
\end{align*}
over beam directions $(\theta_h,\theta_v) \in [-\frac{\pi}{2},\frac{\pi}{2}) \times [-\frac{\pi}{4},\frac{\pi}{4})$.  In Fig. \ref{fig:bp_more}, we plot beam patterns  in each target {beam region} based on the reference gain $\mathrm{G}_{\mathrm{BF}}(\theta_h,\theta_v,\bc_{({\check{q},\check{p})}_{|\theta_h,\theta_v}})$ such that
\begin{align*}
&{(\check{q},\check{p})}_{|\theta_h,\theta_v}=\argmax_{(q ,p) \in \mathcal{Q} }\big|  \bc_{q,p}^H \bd_M\big( \pi  \sin \theta_h \cos \theta_v, \pi \sin \theta_v \big)  \big|^2
\end{align*}
where $\mathcal{Q}=\{1, \cdots, Q_h\} \times \{1, \cdots, Q_v\}$. {As shown in Fig. \ref{fig:bp_one}, the proposed codebook can generate higher reference gains {that will allow} the transmitter {to efficiently sound} mmWave channels as well as to facilitate highly directional data transmission. In Fig. \ref{fig:bp_more}, the mean of the reference gains of the proposed codebook, the proposed codebook considering the guard band, and the codebooks from \cite{Ref_Hur13} and \cite{Ref_Alk14} are $0.553$, $0.467$, $0.397$, and $0.440$, respectively. Furthermore, it is also shown that the proposed codebook in {Section \ref{sec:overlap}} alleviates sharp dips between consecutive beams.  Although the beam pattern may have lower reference gains compared to the  proposed codebook assuming no guard band, the reference gains are much higher than that of the previously reported codebooks in \cite{Ref_Hur13,Ref_Alk14}.}

At this point, we pause to discuss the difference between each of the beam patterns. First, we consider the codebook in \cite{Ref_Hur13}. A beamformer in the codebook of \cite{Ref_Hur13} is designed to maximize a minimum reference gain in each target {beam region}, defined primarily for the corresponding beamformer. {The codebook in \cite{Ref_Hur13} can generate beam patterns with uniform  reference gains, while the value of reference gains are much lower than that of the proposed codebook. {Beam patterns having} uniform beamforming gains are essential for providing  same quality of service to uniformly distributed users, {but} the low reference gains may restrict the data rate performance. Next, we consider the codebook in \cite{Ref_Alk14}. Each beamformer in the proposed codebook is optimized over  beamformer candidates generated using $I^{L_h+L_v-2}$ sets of equal gain vectors  $(\bg_h,\bg_v) \in \cG^{I}_{L_h} \times \cG^{I}_{L_v}$, while the codebook in  \cite{Ref_Alk14} is designed by using a single set of all ones vectors $(\bar{\bg}_h,\bar{\bg}_v)$ in (\ref{eq:single_set}). The codebook in  \cite{Ref_Alk14} is one of several codebook candidates of the proposed algorithm. Therefore, the proposed codebook suppresses the MSE between the ideal beam patterns and the actual beam patterns better than {the beams} in \cite{Ref_Alk14}.}

\subsection{{Data rate performance}}

{We now evaluate the performance of the codebooks based on the expected data rate defined as
\begin{align}
\label{eq:BF}
\mathrm{R}&=\mathrm{E}\Big[ \log_2 \big( 1+\rho|\bh^H\bc |^2 \big)  ~\big|~ \| \bh\|_2^2=M  \Big],
\end{align}
where ${\bc}$ is the selected beamformer for data transmissions.} The preferred beamformer ${\bc}$ in (\ref{eq:BF}) is chosen based on hard-decision beam alignment algorithms \cite{Ref_Hur11,Ref_Hur13,Ref_Son13,Ref_Son14}. %In this paper, we consider single round and multiple rounds beam alignment algorithms.
The data rate performances of the codebooks are evaluated from Monte Carlo simulations with $10,000$ independent channel realizations. For demonstrations, we consider two channel scenarios based on the street geometry conditions under ray-like propagation assumptions \cite{Ref_Zha10}.  In the first scenario, we consider channels consisting of a LOS path and three NLOS paths. The mmWave channel model in \cite{Ref_Hur13, Ref_Son13} is used for simulation. Based on the channel measurements in \cite{Ref_Muh10}, the Ricean K-factor is set to $13.5~\mathrm{dB}$. In the second scenario, the channel vector is characterized {by three NLOS paths without any LOS path \cite{Ref_Akd14,Ref_Zha10}.} We assume that $\|\bh\|_2^2=M$ for fair comparison between two channel scenarios.

\begin{figure*}[!t]
\normalsize
\centering
\subfigure[$(M_h,M_v)=(12,6)$]{\includegraphics[width=0.4825\textwidth]{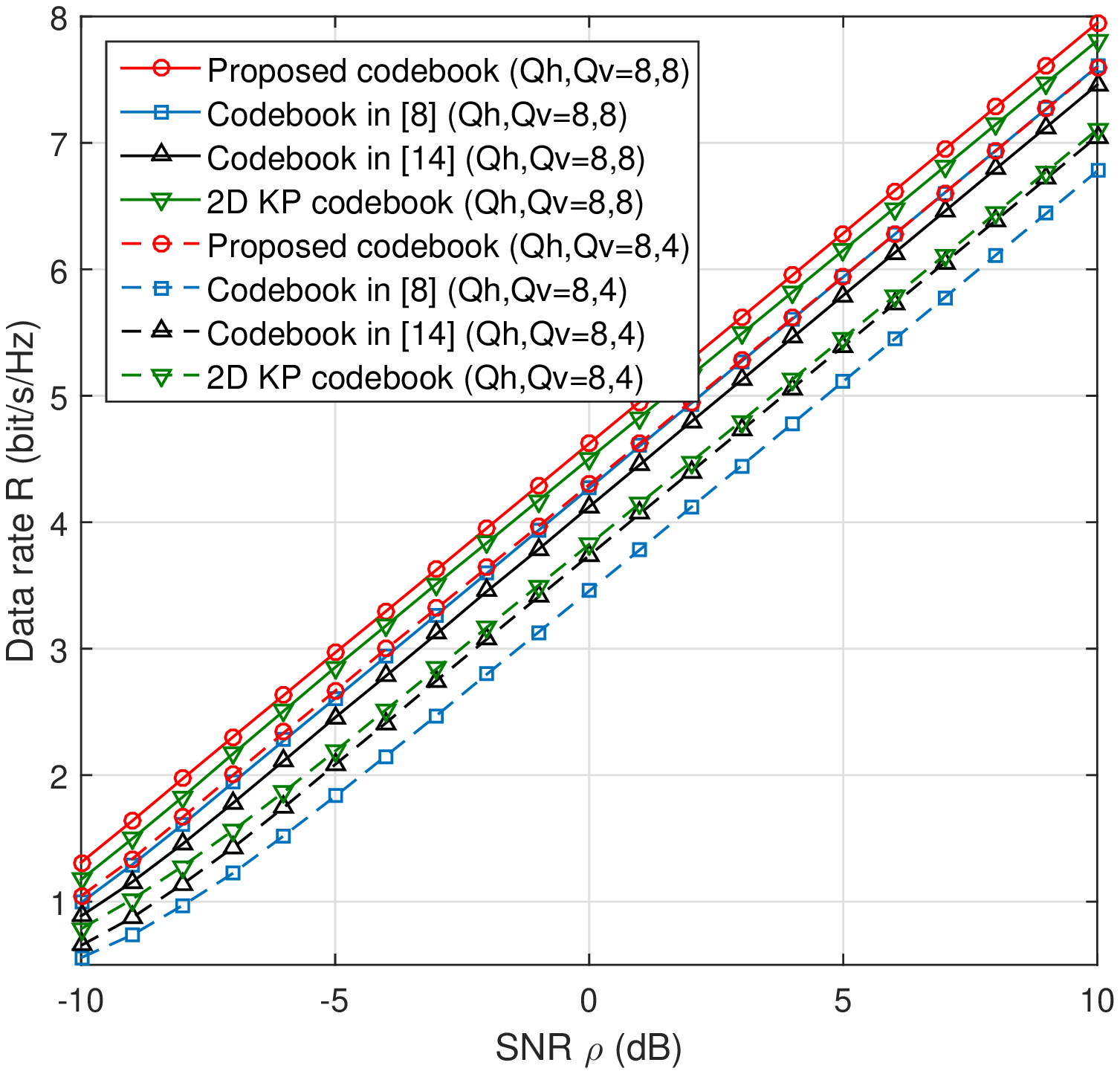}}
\hfil
\subfigure[$(M_h,M_v)=(16,8)$]{\includegraphics[width=0.4825\textwidth]{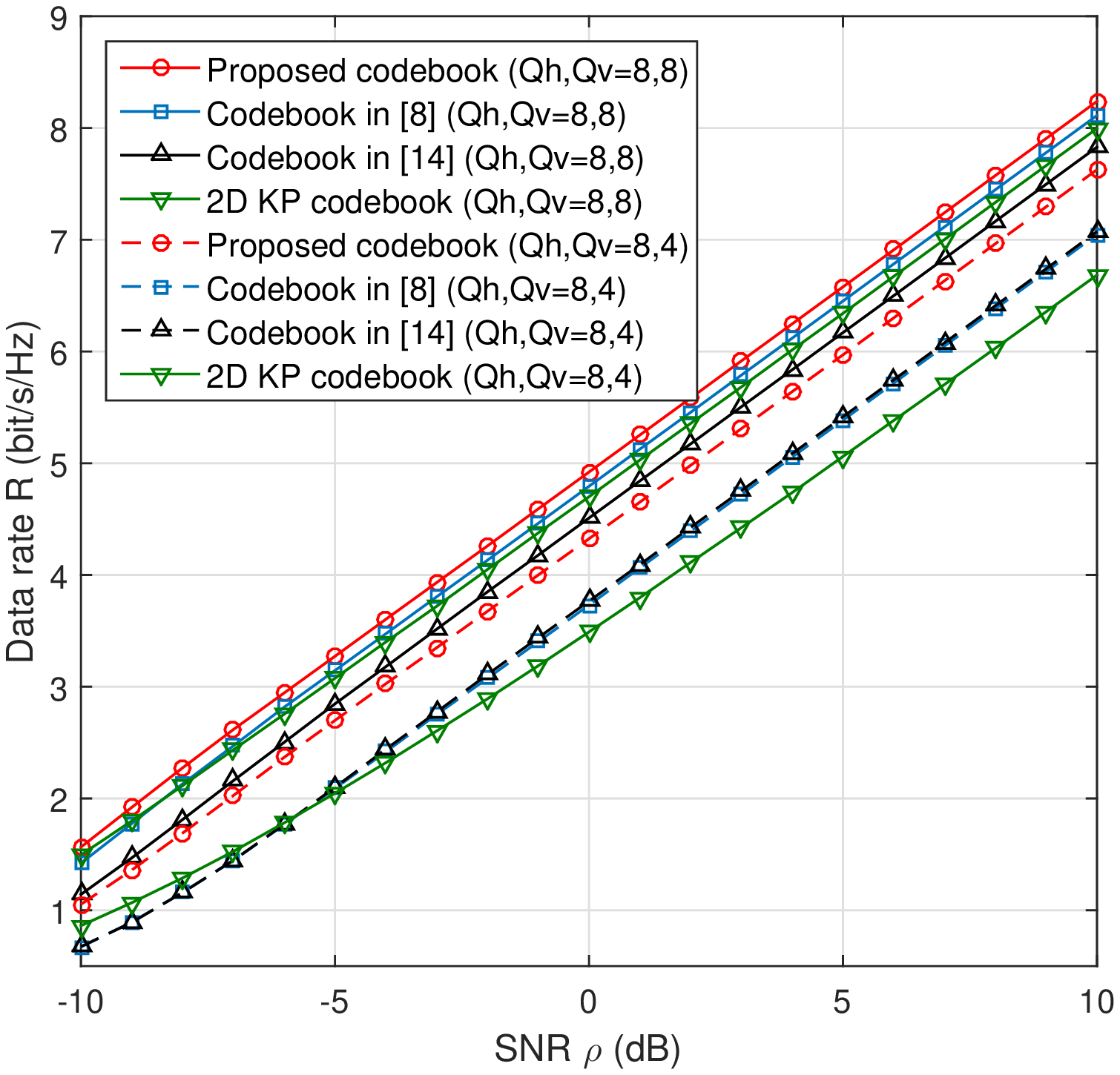}}
\hfil
\subfigure[$(M_h,M_v)=(20,10)$]{\includegraphics[width=0.4825\textwidth]{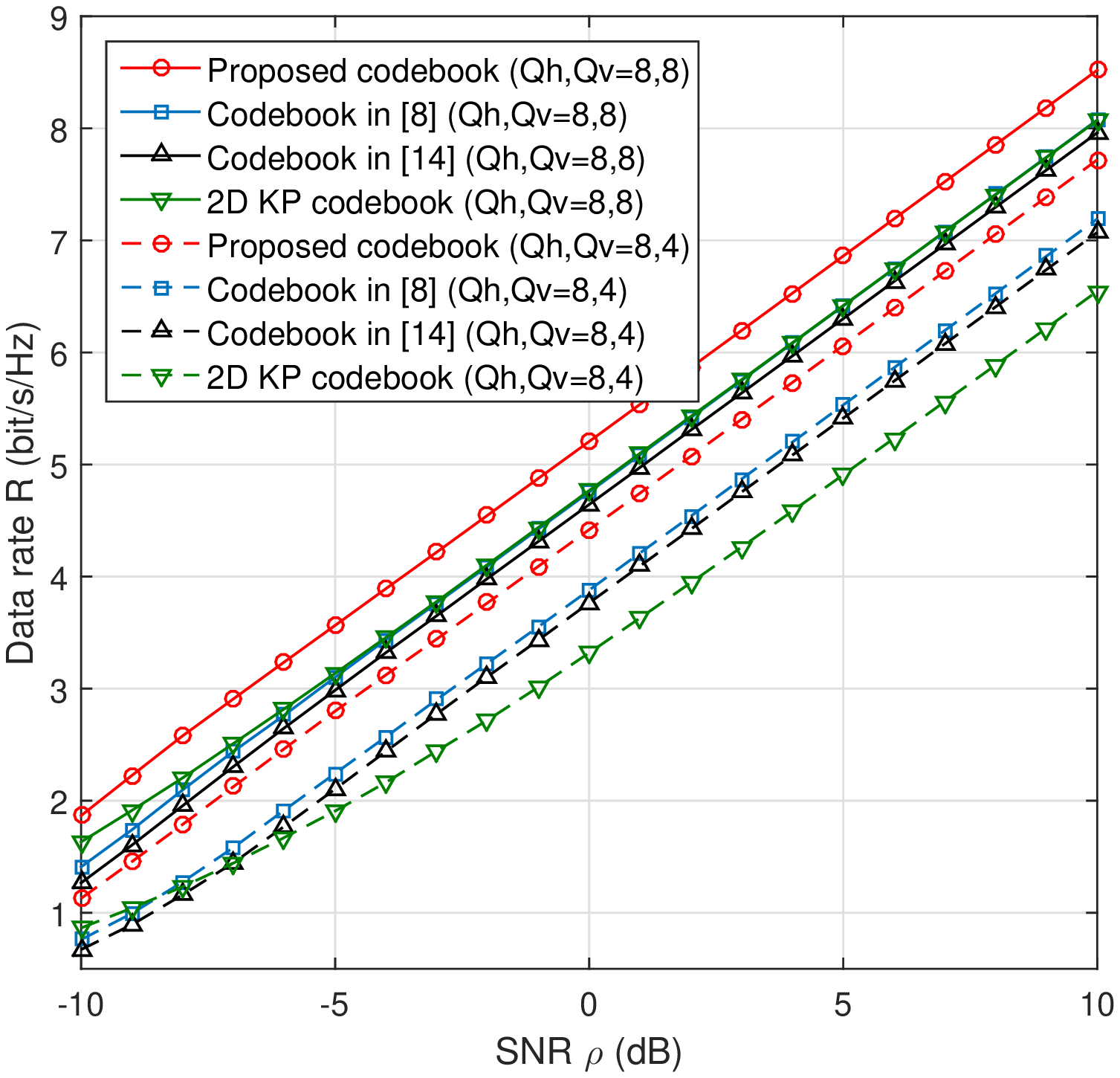}}
\hfil
\subfigure[$(M_h,M_v)=(24,12)$]{\includegraphics[width=0.4825\textwidth]{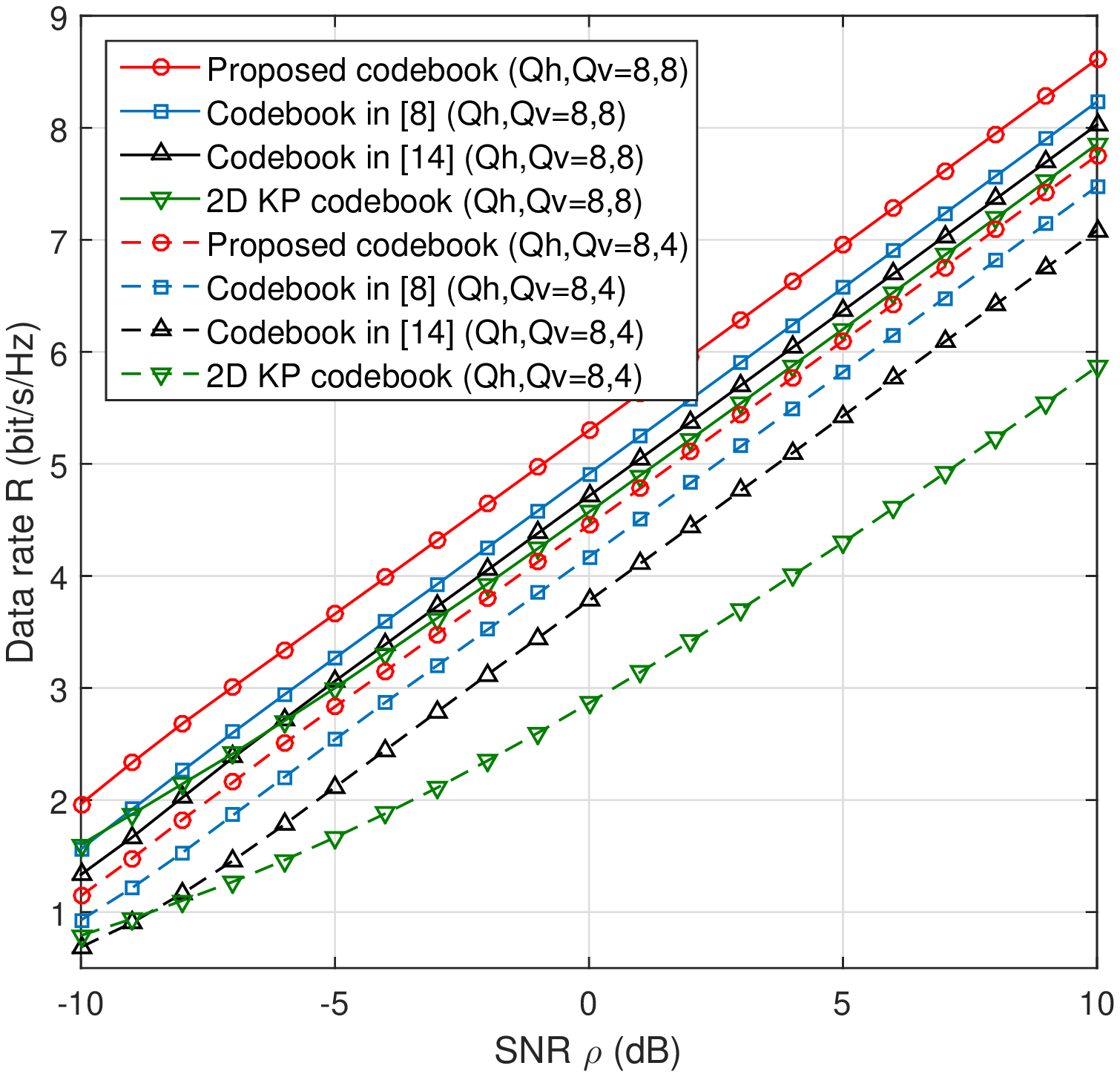}}
\caption{{Data rate based on the channel vectors formed by NLOS paths without any LOS path.}}
\label{fig:bg_single_nlos}
\end{figure*}

We consider a beam alignment approach {for the first channel scenario.} The transmit beamformer is chosen as ${\bc}=\bc_{\hat{q},\hat{p}}$, where $(\hat{q},\hat{p})$ is the index of the selected beamformer in (\ref{eq:beam}). In the beam alignment approach, we compare the ratio between the first and second largest test samples
\begin{align*}
\tau \doteq   {|  \sqrt{\rho}{\bh}^H\bc_{\hat{q},\hat{p}}+n_{\hat{q},\hat{p}}  |^2}/{|  \sqrt{\rho}{\bh}^H\bc_{\dot{q},\dot{p}}+n_{\dot{q},\dot{p}}  |^2}
\end{align*}
with $(\hat{q},\hat{p})$ in (\ref{eq:beam}) and
\begin{align}
\label{eq:secondbeam}
(\dot{q},\dot{p}) = \argmax_{(q,p)  \in \mathcal{Q} \setminus  (\hat{q},\hat{p})   }   {| \sqrt{\rho}\bh^H\bc_{q,p}+n_{q,p} |^2}.
\end{align}
{To avoid beam misalignment, the receiver asks the transmitter to perform an additional cycle of channel sounding if the ratio is smaller than a design parameter $\tau_{t}$. For simulation, the design parameter is set to $\tau_t=2$. In this case, each test sample on two cycles of channel sounding is combined together. The selected transmit beamformer is then communicated to the transmitter via a feedback link employing overheads of $\log_2{Q_h Q_v}$ bits.} {In Fig. \ref{fig:bg_single}, the data rates of different codebooks are compared in UPA structures $(M_h,M_v)=(12,6),~(16,8),~(20,10),~(24,12)$. It is shown that the proposed codebook scans the mmWave channels better and {provides} higher data rates than the codebooks in \cite{Ref_Hur13,Ref_Alk14}, and 2D KP codebook in \cite{Ref_3GPP150560}.}}

\begin{figure*}[!t]
\begin{align}
\nonumber
(\bF^\textrm{opt}_{q,p},\bv^\textrm{opt}_{q,p}) &\stackrel{(a)} =\argmin_{{\bF}, {\bv}} \int_{-\pi}^{\pi}  \int_{-\pi}^{\pi} \big|   G_{1,1}^{\textrm{ideal}} \big(\psi_h-\Delta_{h}^q,\psi_v-\Delta_{v}^p  \big) - G \big(\psi_h,\psi_v, \bF \bv \big)  \big|^2 d\psi_v d\psi_h
\\
\label{eq:16}
&\stackrel{(b)} =\argmin_{{\bF}, {\bv}} \int_{-\pi}^{\pi}  \int_{-\pi}^{\pi} \big|   G_{1,1}^{\textrm{ideal}} \big(\psi_h-\Delta_{h}^q,\psi_v-\Delta_{v}^p  \big) - G \big(\psi_h-\Delta_{h}^q,\psi_v-\Delta_{v}^p, {\mathrm{T}(\bF,-\Delta_{h}^q,-\Delta_{v}^p }) {\bv} \big)  \big|^2 d\psi_v d\psi_h.
\end{align}
\begin{align}
\nonumber
&(\tilde{\bF}^\textrm{opt}_{q,p},\bv^\textrm{opt}_{q,p})=\argmin_{\tilde{\bF}, {\bv}} \int_{-\pi - \Delta_{h}^q}^{\pi - \Delta_{h}^q }  \int_{-\pi - \Delta_{v}^p }^{\pi - \Delta_{v}^p} \big|   G_{1,1}^{\textrm{ideal}}\big(\phi_h,\phi_v\big) - G\big(\phi_h,\phi_v, {\tilde{\bF}}{\bv}\big)  \big|^2 d\phi_v d\phi_h
\\
\nonumber
&=\argmin_{\tilde{\bF}, {\bv}} \int_{-\pi - \Delta_{h}^q}^{\pi - \Delta_{h}^q }  \bigg[ \int_{-\pi - \Delta_{v}^p}^{- \pi} \big|   G_{1,1}^{\textrm{ideal}}\big(\phi_h,\phi_v\big) - G\big(\phi_h,\phi_v, {\tilde{\bF}}{\bv}\big)  \big|^2 d\phi_v +\int_{-\pi   }^{\pi - \Delta_{v}^p } \big|   G_{1,1}^{\textrm{ideal}}\big(\phi_h,\phi_v\big) - G\big(\phi_h,\phi_v, {\tilde{\bF}}{\bv}\big)  \big|^2 d\phi_v\bigg] d\phi_h
\\
\nonumber
&\stackrel{(a)} =\argmin_{\tilde{\bF}, {\bv}} \int_{-\pi - \Delta_{h}^q}^{\pi - \Delta_{h}^q }  \bigg[ \int_{ \pi - \Delta_{v}^p}^{  \pi} \big|   G_{1,1}^{\textrm{ideal}}\big(\phi_h,\phi_v\big) - G\big(\phi_h,\phi_v, {\tilde{\bF}}{\bv}\big)  \big|^2 d\phi_v +\int_{-\pi   }^{\pi - \Delta_{v}^p } \big|   G_{1,1}^{\textrm{ideal}}\big(\phi_h,\phi_v\big) - G\big(\phi_h,\phi_v, {\tilde{\bF}}{\bv}\big)  \big|^2 d\phi_v\bigg] d\phi_h
\\
\label{eq:17}
&=\argmin_{\tilde{\bF}, {\bv}} \int_{-\pi}^{\pi  }  \int_{ -\pi }^{  \pi} \big|   G_{1,1}^{\textrm{ideal}}\big(\phi_h,\phi_v\big) - G\big(\phi_h,\phi_v, {\tilde{\bF}}{\bv}\big)  \big|^2 d\phi_v  d\phi_h =({\bF}^\textrm{opt}_{1,1},\bv^\textrm{opt}_{1,1}).
\end{align}
\begin{align}
\nonumber
G &\big(\psi_h,\psi_v, \bF \bv \big)\stackrel{(a)}=\Big| \big( \bd_M(\psi_h+\Delta_{h}^q,\psi_v+\Delta_{v}^p)  \odot \tilde{\bd}_M(-\Delta_{h}^q,-\Delta_{v}^p) \big)^H \big( \bF \odot  \mathbf{1}_{M,N} \big)\bv  \Big|^2
\\
\nonumber
\stackrel{(b)}=&\Big| \big( \bd_M(\psi_h+\Delta_{h}^q,\psi_v+\Delta_{v}^p)  \odot \tilde{\bd}_M(-\Delta_{h}^q,-\Delta_{v}^p)\big)^H\big( \bF  \odot \big( \big( \tilde{\bd}_M(\Delta_{h}^q,\Delta_{v}^p) \mathbf{1}_{1,N}  \big) \odot \big( \tilde{\bd}_M(-\Delta_{h}^q,-\Delta_{v}^p) \mathbf{1}_{1,N} \big) \big) \big) \bv  \Big|^2
\\
\nonumber
\stackrel{(c)}=&\Big| \big( \bd_M(\psi_h+\Delta_{h}^q,\psi_v+\Delta_{v}^p)  \odot \tilde{\bd}_M(-\Delta_{h}^q,-\Delta_{v}^p) \big)^H \big( \mathrm{T}\big(\bF , \Delta_{h}^q,\Delta_{v}^p \big)  \odot \tilde{\bd}_M(-\Delta_{h}^q,-\Delta_{v}^p) \mathbf{1}_{1,N}\big) \bv  \Big|^2
\\
\nonumber
=&\Big| \Big( \big(\bd_M(\psi_h+\Delta_{h}^q,\psi_v+\Delta_{v}^p)  \odot \tilde{\bd}_M(-\Delta_{h}^q,-\Delta_{p})\big)^H
\\
\nonumber
&~~~~~~~~~~~~~~~~~~~~~~~~~~~~~~~~~~\Big[ \mathrm{T}\big(\bF , \Delta_{h}^q,\Delta_{v}^p \big)_{:,1} \odot \tilde{\bd}_M(-\Delta_{h}^q,-\Delta_{v}^p), \cdots,  \mathrm{T}\big(\bF , \Delta_{h}^q,\Delta_{v}^p \big)_{:,N} \odot \tilde{\bd}_M(-\Delta_{h}^q,-\Delta_{v}^p) \Big] \Big) \bv  \Big|^2
\\
\label{eq:pro}
\stackrel{(d)}=&\Big|  \bd^H_M(\psi_h+\Delta_{h}^q,\psi_v+\Delta_{v}^p)   \mathrm{T}\big(\bF , \Delta_{h}^q,\Delta_{v}^p \big) \bv  \Big|^2=G  \big(\psi_h + \Delta_{h}^q ,\psi_v + \Delta_{v}^p, \mathrm{T}\big(\bF , \Delta_{h}^q,\Delta_{v}^p \big) \bv \big).
\end{align}
\hrulefill
\end{figure*}

{We also evaluate data rate performance considering a second channel scenario consisting of three NLOS paths.\footnote{Although we choose a single beamformer, we could select and combine multiple beamformers as in \cite{Ref_Cho15,Ref_Son16_con} to fully support channels consisting of multiple NLOS paths.}} {In Fig. \ref{fig:bg_single_nlos}, the data rate based on different codebooks are compared in UPA structures, i.e., $(M_h,M_v)=(12,6),~(16,8),~(20,10),~(24,12)$. In the case of the NLOS channel scenario, it is shown that the proposed codebooks select a dominant NLOS path better than the previously reported codebooks \cite{Ref_Hur13,Ref_Alk14}, and 2D KP codebook in \cite{Ref_3GPP150560}.}

\section{Conclusions}
In this paper, we proposed a beam pattern design algorithm suited to the directional characteristics of the mmWave channels corresponding to UPAs. We proposed an iterative algorithm to construct small-sized beam alignment codebooks for mmWave systems. A hybrid beamforming architectures using a mixture of analog and digital beamforming {was} considered to design effective beams suitable to large-scale mmWave systems with limited RF chains. In the proposed  algorithm, each beamformer is constructed to minimize the MSE between its actual beam pattern and the corresponding ideal beam pattern. We developed a simplified approach to solve the approximated MSE minimization problem. In addition, we used the OMP algorithm to design beamformers satisfying the hybrid beamforming setup. {The data rate performance of proposed codebooks is verified through Monte Carlo simulations. Numerical results show that our codebooks outperform previously reported codebooks in mmWave channels using UPA structures.}

%%%%%%%%%%%%%%%%%%%%%%%%%%%%%%%%%%%%%%%%%%%%%%
%%                                          %%
%%                 Appendix                 %%
%%                                          %%
%%%%%%%%%%%%%%%%%%%%%%%%%%%%%%%%%%%%%%%%%%%%%%

\appendices

\section{Proof of Lemma \ref{lm:03}}
\label{sec:A}

\begin{figure}[!t]
\centering
{\includegraphics[width=0.435\textwidth]{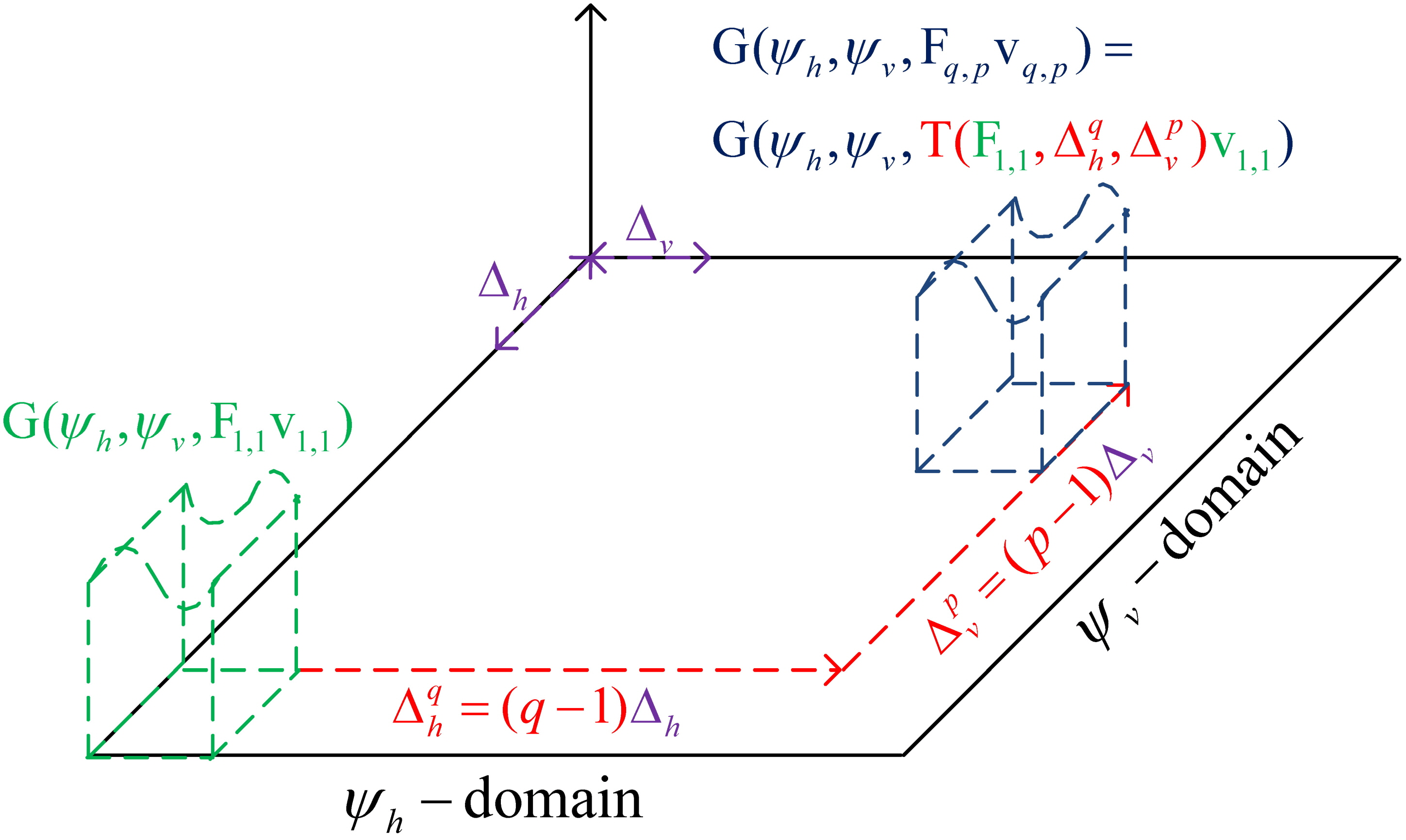}}
\caption{{Shifted beam pattern from an optimized beamformer.}}
\label{fig:shift}
\end{figure}

\begin{prop}
\label{prop:ibp}
Any ideal beam pattern is defined  by shifting the $(1,1)$-th ideal beam pattern according to
\begin{align*}
G_{q,p}^{\textrm{ideal}}(\psi_h,\psi_v)=G_{1,1}^{\textrm{ideal}}(\psi_h-\Delta_{h}^q,\psi_v-\Delta_{v}^p).
\end{align*}
\end{prop}
{Also, the directions of the actual beam pattern is also shifted  based on the following lemma, as depicted in Fig. \ref{fig:shift}.}
\begin{lemma}
\label{lm:04}
The reference gain is rewritten by using the phase shifting function $\mathrm{T}\big(\bF , \Delta_{h}^q,\Delta_{v}^p \big)$ {in (\ref{eq:ps}) such as}
\begin{align*}
&G \big(\psi_h,\psi_v, \bF \bv \big)=G  \big(\psi_h + \Delta_{h}^q ,\psi_v+  \Delta_{v}^p, \mathrm{T}\big(\bF , \Delta_{h}^q,\Delta_{v}^p \big) \bv \big).
\end{align*}
{Please see} Appendix \ref{sec:B} for the proof.
\end{lemma}

The  problem for the $(q,p)$-th beamformer is rewritten in (\ref{eq:16}). Note that $(a)$ is derived based on Proposition \ref{prop:ibp} and $(b)$ is derived based on Lemma \ref{lm:04}. Then, we modify the rewritten minimization problem in (\ref{eq:16}) to discuss the link with the solution for $(1,1)$-th beamformer $\big({\bF}^\textrm{opt}_{1,1},\bv^\textrm{opt}_{1,1}\big)$. First,  variables for the integration in (\ref{eq:16}) are changed by defining new variables as $\phi_a \doteq \psi_a-\Delta_{a}^b$ in each domain of the double integral. We next change a dummy variable by defining the phase shifted matrix as
\begin{align*}
\tilde{\bF} \doteq {\mathrm{T}\big( \bF,-\Delta_{h}^q ,-\Delta_{v}^p \big).}
\end{align*}
By plugging redefined variables $\phi_h$, $\phi_v$, and  $\tilde{\bF}$ into (\ref{eq:16}), the minimization problem is rewritten as in (\ref{eq:17}). Note that $(a)$ is derived because the ideal beam pattern and the reference gain are periodic functions with a period of $2\pi$ over  $\phi_h$ and $\phi_v$.

{The solution {to the} MSE problem in  (\ref{eq:17}) is identical to that for the $(1,1)$-th beamformer. Therefore, it is clear that
\begin{align*}
\tilde{\bF}^\textrm{opt}_{q,p} ={\bF}^\textrm{opt}_{1,1},~~~\bv^\textrm{opt}_{q,p}=\bv^\textrm{opt}_{1,1},
\end{align*}
where $\tilde{\bF}^\textrm{opt}_{q,p}=\mathrm{T} \big( \bF^\textrm{opt}_{q,p} , -\Delta_{h}^q, -\Delta_{v}^p\big)$ is rewritten by definition of the phase shifted matrix $\tilde{\bF}$.} Because the phase shifting function in  (\ref{eq:ps}) is reversible, the relationship is rewritten as
\begin{align*}
&\mathrm{T} \big( \bF^\textrm{opt}_{1,1}, \Delta_{h}^q, \Delta_{v}^p\big)= \mathrm{T} \big( \mathrm{T} \big( \bF^\textrm{opt}_{q,p} , -\Delta_{h}^q, -\Delta_{v}^p \big) , \Delta_{h}^q, \Delta_{v}^p\big)
\\
&=\bF^\textrm{opt}_{q,p} \odot \big( \tilde{\bd}_M(-\Delta_{h}^q,-\Delta_{v}^p)\odot\tilde{\bd}_M(\Delta_{h}^q,\Delta_{v}^p)   \big)   \mathbf{1}_{1,N}
\\
&=\bF^\textrm{opt}_{q,p} \odot  \mathbf{1}_{M,N}=\bF^\textrm{opt}_{q,p}.
\end{align*}
Finally, the analog beamsteering matrix and baseband beamformer for the $(q,p)$-th  beamformer can be derived  as
\begin{align*}
\bF^\textrm{opt}_{q,p} =   \mathrm{T} \big( \bF^\textrm{opt}_{1,1}, \Delta_{h}^q,\Delta_{v}^p\big),~~~\bv^\textrm{opt}_{q,p} =\bv^\textrm{opt}_{1,1}.
\end{align*}

\section{Proof of Lemma \ref{lm:04}}
\label{sec:B}
{The reference gain $G \big(\psi_h,\psi_v, \bF \bv \big)$ is rewritten by using the phase shifting functions $\mathrm{T}\big(\bF , \Delta_{h}^q,\Delta_{v}^p \big)$ in (\ref{eq:pro}). {Note that $(a)$ of (\ref{eq:pro})} is derived based on the property of array vectors
\begin{align*}
&\bd_M(\psi_{1}+\psi_{3},\psi_{2}+\psi_{4})=\bd_M(\psi_{1},\psi_{2}) \odot \tilde{\bd}_M(\psi_{3},\psi_{4}),
%\\
%&\tilde{\bd}(\psi_{a},\psi_{b})=\frac{{\bd}(\psi_{a},\psi_{b})}{\|{\bd}(\psi_{a},\psi_{b}) \odot {\bd}(\psi_{a},\psi_{b}) \|_2},
\end{align*}
$(b)$ is derived because $\mathbf{1}_{M,N}$ is decomposed as
\begin{align*}
\mathbf{1}_{M,N}&=\big(\tilde{\bd}_M(\psi_{1},\psi_{2}) \odot \tilde{\bd}_M(-\psi_{1},-\psi_{2})\big)\mathbf{1}_{1,N}
\\
&=\big( \tilde{\bd}_M(\psi_{1},\psi_{2})\mathbf{1}_{1,N} \big) \odot \big( \tilde{\bd}_M(-\psi_{1},-\psi_{2})\mathbf{1}_{1,N} \big),
\end{align*}
and $(c)$ is rewritten by the definition of the phase shifting function $\mathrm{T}\big(\bF , \Delta_{h}^q ,\Delta_{v}^p \big)$ in Lemma \ref{lm:03} and the associative property of the Hadamard product
\begin{align*}
\bA \odot (\bB \odot \bC)=(\bA \odot \bB) \odot \bC
\end{align*}
for arbitrary matrices of the same size. Finally, $(d)$ is derived based on the formulation  between vectors  $\ba=[a_1,\cdots,a_M]^T$, $\bb=[b_1,\cdots,b_M]^T$, and $\bc=[c_1,\cdots,c_M]^T$,
\begin{align*}
(\ba \odot \bb)^H(\bc \odot \bb)&=\sum_{m=1}^{M}  a_{m}^* c_{m}|b_{m}|^2
\\
&=\ba^H\bc,
\end{align*}
which holds with the condition of $| b_{m}|^2=1$ for all $m \in \{1,\cdots, M\}$. Note  that $(d)$ satisfies this condition  because each element of $\tilde{\bd}_M(-\Delta_{h}^q,-\Delta_{v}^p)$ has a unit gain, i.e., $\big| \big(\tilde{\bd}_M(-\Delta_{h}^q,-\Delta_{v}^p)\big)_{m} \big|^2=1$ for all $m \in \{1,\cdots, M\}$.
}

\bibliographystyle{IEEEtran}
\bibliography{refs_HC}

\end{document}